\documentclass[10pt,preprint2]{aastex}
\newcommand{\ifpp}[1]{#1}
\newcommand{\ifms}[1]{}
\newcommand{\etal}{{\rm et~al.~}}
\def\ltsima{$\; \buildrel < \over \sim \;$}
\def\simlta{\lower.5ex\hbox{\ltsima}} % < over ~
\newcommand{\simlt}{\ifmmode {\simlta\ } \else {$\simlta$\ } \fi}
\newcommand{\degree}{\ifmmode {^{\circ}} \else {$^{\circ}$} \fi}
\newcommand{\degrees}{\ifmmode {^{\circ}} \else {$^{\circ}$} \fi}
\newcommand{\ie}{{\em i.e.}}
\newcommand{\unit}[1]{\ifmmode {\rm\ #1\,} \else {$\rm #1$} \fi}
\newcommand{\quarter}{\ifmmode {\frac{1}{4}} \else {$\frac{1}{4}$} \fi}

\newcommand{\tten}[1]{\ifmmode {\times 10^{#1}} \else {$\times 
10^{#1}$} \fi}
\newcommand{\tentothe}[1]{\ifmmode {10^{#1}} \else {$10^{#1}$} \fi}

\newcommand{\doublet}{\ifmmode {\lambda\lambda} \else 
{$\lambda\lambda$} \fi}
\newcommand{\singlet}{\ifmmode {\lambda} \else {$\lambda$} \fi}

\newcommand{\percmcubed}{\unit{cm^{-3}}}

\newcommand{\captionone}{
(a) - The azimuthally averaged radial 
profile of the total 0.13 - 0.18 keV EUV count rate (solid), the 
EUV count rate in this band produced by the X-ray plasma  plus the 
EUV background count rate (dashed), and the EUV background alone 
(dotted).  (b) - The azimuthally 
averaged radial profile of the 0.13 - 0.18 keV EUV 
excess count rate in the Coma Cluster.
}
\newcommand{\captiontwo}{The ratio of the azimuthally 
averaged EUV excess flux/ROSAT 0.5-2.4 keV X-ray flux as a 
function of increasing distance from the 
cluster center.}
\newcommand{\captionthree}{
Sky map of the EUV excess in the 
Coma Cluster in 1$^\prime$ square bins (J2000). The 
isophotal lines shown are at 90\%, 50\% and 
23\% of the peak EUV excess emission.}
\newcommand{\captionfour}{
X-ray emission in the Coma Cluster derived from ROSAT PSPC 
data.}
\newcommand{\captionfive}{
The correlation coefficients of two independent data sets of the 
EUV excess in Coma as a function of the size of the sky map binning is 
shown by diamonds. The correlation of the EUV excess with the X-ray 
emission as a function of the size of the sky map binning is shown by
triangles.}
\newcommand{\captionsix}{
A plot showing observations constraining the secondary
emission model. The solid lines are a schematic
representation of the radio observations with the
range of spectral indices justified by Thierbach et
al. 2003. The solid squares are observed values of the
EUV and HRX fluxes. The open square is the
observational upper limit to the gamma-ray flux. Our
model produces the observed EUV flux while not
exceeding the radio and HRX fluxes for a range of
reasonable magnetic fields. The open triangle shows
the $\gamma$-ray flux produced by our model.}

\begin{document}
\shortauthors{Bowyer, \etal}
\shorttitle{The Source of EUV Emission in the Coma Cluster}
\title{
The EUV Emission in the Coma Cluster of Galaxies and 
the Underlying Source of this Radiation
} 

\author{Stuart Bowyer, Eric J. Korpela \& Michael Lampton}
\affil{Space Sciences Laboratory, Univ. of Calif.}
\author{T.~W. Jones}
\affil{Astronomy Dept. Univ. of Minn.}

\begin{abstract}
 Observations with the Extreme Ultraviolet Explorer (EUVE) have shown the Coma Cluster to
 be a source of EUV emission in excess of that
produced
 by X-ray gas in the cluster. We have re-examined the
EUVE
data on this cluster in an attempt to obtain clues as
 to the origin of this emission. We find two important
 new results. First, the ratio between the azimuthally
 averaged EUV excess emission and the ROSAT hard X-ray flux is
 constant as a function of distance from the cluster center outward. 
Second, a correlation analysis
between
 the EUV excess emission and the X-ray emission shows that
 on a detailed level
the
 EUV excess is spatially closely related to the
X-ray
 emission. These findings contradict 
previous
 suggestions as to the underlying source of the
diffuse
 EUV emission in Coma and provide important information in regards to the
 true source of this emission.  We propose a new explanation for the source of this 
emission: inverse Compton scattering of microwave background photons by 
secondary electrons and positrons.  
We explore this possibility in some detail 
and show that
 it is consistent with all of the available observational evidence.
The parent cosmic ray protons may 
have been produced by any of a number of sources, including supernovae, active
galaxies, galactic winds, and 
cluster formation shocks, but we believe that the most likely source is 
cluster formation shocks. 
 If the EUV emission
 in the Coma Cluster is, in fact, the result of secondary electrons, this may be
 the
 only direct
 evidence for secondary electrons in the intracluster medium of a 
cluster of galaxies, since recent work suggests that secondary
electrons may not be the cause of radio halos. \\\ \\
\end{abstract}

\section{Introduction}

Observations with the Extreme Ultraviolet Explorer (EUVE) provided evidence 
that a number of clusters of galaxies emit
excess EUV emission in the cores of the clusters. The first clusters 
reported to have EUV excesses were the 
Virgo cluster (Lieu \etal 1996a; Bowyer \etal 1996) and the 
Coma Cluster (Lieu \etal 1996b).  Thereafter 
EUV emission was reported for Abell 1795 
(Mittaz, Lieu \& Lockman, 1998) and Abell 
2199 (Lieu \etal 1999a). 
These early works 
employed a variety of data analysis schemes 
that were later found to be incorrect 
(Bowyer, Bergh\"ofer, \& Korpela, 1999), 
primarily because incorrect methods were 
used to account for the sensitivity profile, or 
exposure map, of the telescope.  The only clusters that have been 
determined to
have an EUV excess using uncontested data analysis procedures are the 
Virgo
cluster (Bergh\"ofer \etal 2000) and the Coma Cluster (Bowyer \etal 
1999).

Subsequent to the analysis of Bowyer \etal (1999) additional EUV data on the 
Coma Cluster were obtained with EUVE. In this paper we re-examine the 
excess EUV emission in the Coma Cluster using all the EUVE data 
available on this cluster. 
We obtain important new information on the character of the EUV 
emission in this cluster. Given 
these
new results, we 
provide strong evidence that the EUV emission is produced by secondary 
electrons and positrons in the intracluster medium (ICM).  This 
finding
may well be the only secure evidence of the presence of secondary 
electrons and positrons in an intracluster medium.

\section{Data and Data Analysis}

All of the data employed were obtained with the Deep Survey (DS) 
telescope of EUVE (Bowyer \& Malina 1991). In 
Table~1 we provide an observing log of the observations.  The total observing 
time was
390 ks.
\begin{table}
\begin{center}
\begin{tabular}{lr}
\hline
Date    & \multicolumn{1}{c}{Duration (ks)} \\
\hline
\hline
12/25/95-12/28/95	&	50 \\
06/11/96-06/12/96	&	39 \\
01/12/99-01/14/99	&	53 \\
02/04/99-02/07/99	&	76 \\
03/15/99-03/21/99	&	172 \\
\hline
\end{tabular}
\caption{Log of Observations}
\end{center}
\end{table}

Various authors have used a number of data reduction procedures in 
searches for EUV emission from clusters. Because of the misconceptions 
created by 
the use of incorrect analysis procedures, we 
describe the data reduction approach used 
here in some detail. These procedures 
were developed and documented in Bowyer 
\etal 1999. The validity of this approach was 
examined and tested by Bergh\"ofer, 
Bowyer, \& Korpela 2000 and its appropriateness verified. 

First, the Coma data sets were screened to 
exclude noisy data.  The pulse height 
distribution of each set was then examined 
and low energy counts produced by random 
noise were excluded by rejecting counts 
below a low energy threshold.  Since a low 
energy threshold is applied to the data by the 
onboard satellite data processing system, 
this step was not crucial.  Indeed, 
Bergh\"ofer \etal 2000 have shown that 
changing the low energy threshold by as 
much as a factor of two has no effect on the end result.  Nonetheless, 
this approach can improve the quality of the data set in at least some 
cases.  Cosmic rays interacting with 
the spacecraft and the detector produce a 
few high energy counts in the data which we 
removed by upper level thresholding.  These 
counts are only a small fraction of the total 
data set and ignoring this step does not 
significantly affect the end result. 
Nonetheless, these counts were easy to 
remove and we did so.  Corrections were 
then made to account for telemetry 
limitations and detector dead time effects on 
the total observing time; these were $\sim 10\%$.

The next step in our analysis is quite 
important.  A background was obtained 
from regions of the detector that do not view 
photons from the sky.  This background 
arises from energetic charged particles interacting 
with the satellite; these produce charged 
particles within the instrument that trigger 
counts in the detector.  This background 
varies over time scales of weeks to months 
and depends upon geophysical conditions.  
Bergh\"ofer \etal 2000 have shown that this 
background differs by only a factor of two 
over the course of the EUVE mission, but 
given the low counting rates from clusters of 
galaxies it is important that this background 
level be identified in order to establish the 
zero level for each particular observation. 
Accordingly, we established this background independently for each of 
the data sets we employed.  

Most importantly, the correct telescope 
sensitivity profile, or exposure map, was 
used in connection with the analysis of the 
data.  We note similar corrections for the 
instrument sensitivity over the field of view 
are routinely applied in the reduction of 
most observations of diffuse X-ray emission.  
For example, observations of diffuse sources 
with the ROSAT PSPC are routinely 
corrected using an effective area exposure 
map (Snowden \etal 1994). Bowyer \etal 
1999 have provided a map of the EUVE DS 
sensitivity profile using 363 ks of data from 
a variety of blank fields.  The use of a 
sensitivity profile composed of a large 
number of individual blank field data sets 
could, in principle, be questioned.  Indeed, 
Lieu \etal 1999a claimed the EUVE DS 
sensitivity profile varies with time, but no 
analysis validating this claim was provided.  
Bergh\"ofer \etal 2000 carried out a 
detailed investigation of this possibility.  
They compared the 363 ks data set referred 
to above with an assemblage of 425 ks of 
data from a different set of blank fields 
obtained at different times.  The two data 
sets were correlated at the 97\% level, 
consistent with the statistical uncertainties in 
the counts in the individual cells in the two 
data sets. This demonstrated the stability of 
the EUVE DS telescope's sensitivity profile 
over time scales of years.  In our work on 
the Coma Cluster we used a sensitivity 
profile composed of 788 ks of data obtained 
by combining the two blank field data sets described above. 

Because of the different orientations of each 
of the different Coma observations, it was 
necessary to carry out the above steps on 
each of the individual data sets separately. 
The results of each observation were then 
summed. This required a knowledge of the absolute pointing of the 
spacecraft. Because there are no obvious point 
sources that are present in all of the EUV 
images, it is non-trivial to confirm the 
pointing coordinates provided by the 
satellite. A comparison of the location of the 
maximum of the cluster emission in the images shows the relative 
pointing error in the nominal spacecraft 
pointing to be $\sim
0\arcmin.28$.
Since this uncertainty 
is \simlt the estimated point spread function of the 
telescope, we simply added the images using 
the nominal spacecraft pointing. We note, 
however, that any conclusions based on the 
EUVE data will be uncertain at this, or smaller, scales.

Next, the effects of absorption by the 
Galactic interstellar medium (ISM) on the EUV 
flux were determined.  There are a number 
of programs available to determine the 
effects of the ISM on the X-ray flux from 
Galactic and extragalactic sources and any 
of these will provide a result that is 
essentially valid in the X-ray regime.  
However in the EUV, the situation is 
entirely different.  In this band absorption is 
due only to hydrogen, neutral helium, and 
singly ionized helium.  Metals can be 
ignored because they produce insignificant 
absorption in comparison to these species, 
and the reduction of He I and He II due to the
presence of He III can be ignored because there is 
virtually no He III in the ISM
(Heiles \etal 1996). The 
appropriate EUV cross sections must be 
used for H I, He I, and He II, and equally 
importantly, correct columns are needed for 
each of these components.  In particular, the 
amount of H II in the line of sight must be 
established in order to determine the true He 
I and He II columns.  A full discussion of 
these issues and a comparison of the 
differing outcomes with the use of different 
compilations of cross-sections are provided 
in Bowyer \etal 1999.  In this case we used a 
hydrogen column of $8.95\times{10^{19}}$ cm$^{-2}$
(Dickey \& Lockman 1990) with ionization fractions 
and cross sections for Galactic ISM 
absorption as described in detail in Bowyer 
\etal 1999.  We note that Bregman \etal 2003
have shown that small scale variations in the
Galactic ISM can be as large as factor of 3 in
some 1 degree fields containing clusters of galaxies
and this can affect the magnitude of the
EUV excess in these cases.  However, this is not a factor
in regard to the Coma Cluster where there is near spatial
uniformity of the Galactic H{\sc i} column as manifested in the
NRAO map of this region with a spatial resolution of 21\arcmin,
and the finer scale IRAS 100 $\mu$m map.

We then derived the EUV emission 
produced by the high temperature X-ray 
emitting gas using Coma ROSAT PSPC 
archival data. We used a temperature of 9 
keV (Briel et al. 1992). We note that a variety of
temperatures, typically varying from 8 to 9 keV have
been reported for the thermal gas in Coma by various authors.
This variation has only a small effect on the ratio of the X-ray
to EUV flux.   This ratio for an 8 keV plasma is
within 10\% of that for a 9 keV plasma.
We used a factor of 128 to convert counts in the 0.5-2.4 keV
band of the ROSAT PSPC to the EUVE DS-band counts. The PSPC conversion
factor was derived
from the MEKAL plasma code with abundances of 0.3 solar and a 
temperature
of 9 keV.  
We corrected for the Galactic ISM as described 
above.

The next task was to align the X-ray and 
EUV images. A source well away from the 
cluster center was detected at the same sky 
location in both the EUVE DS image and 
the ROSAT soft X-ray image. In both of 
these images the source was $\simlt$  the point-
spread functions of the respective detectors.  
A UV source, A 2305, is located within the 
central portion of the point-spread functions 
in both the EUVE and the ROSAT images.  
A QSO would typically produce a UV, 
EUV, and soft X-ray signature of this 
character. With this source as a fiducial, the 
images were aligned to  $\simlt 0\arcmin.43$. We note that any 
comparisons between the 
EUV and X-ray data are uncertain at, or less 
than, scales of 0\arcmin.43.  We then subtracted the EUV emission due 
to the X-ray gas from 
the total EUV emission detected by EUVE. 

\ifpp{
\begin{figure}[tb]
\begin{center}
\epsscale{2.0}
\plottwo{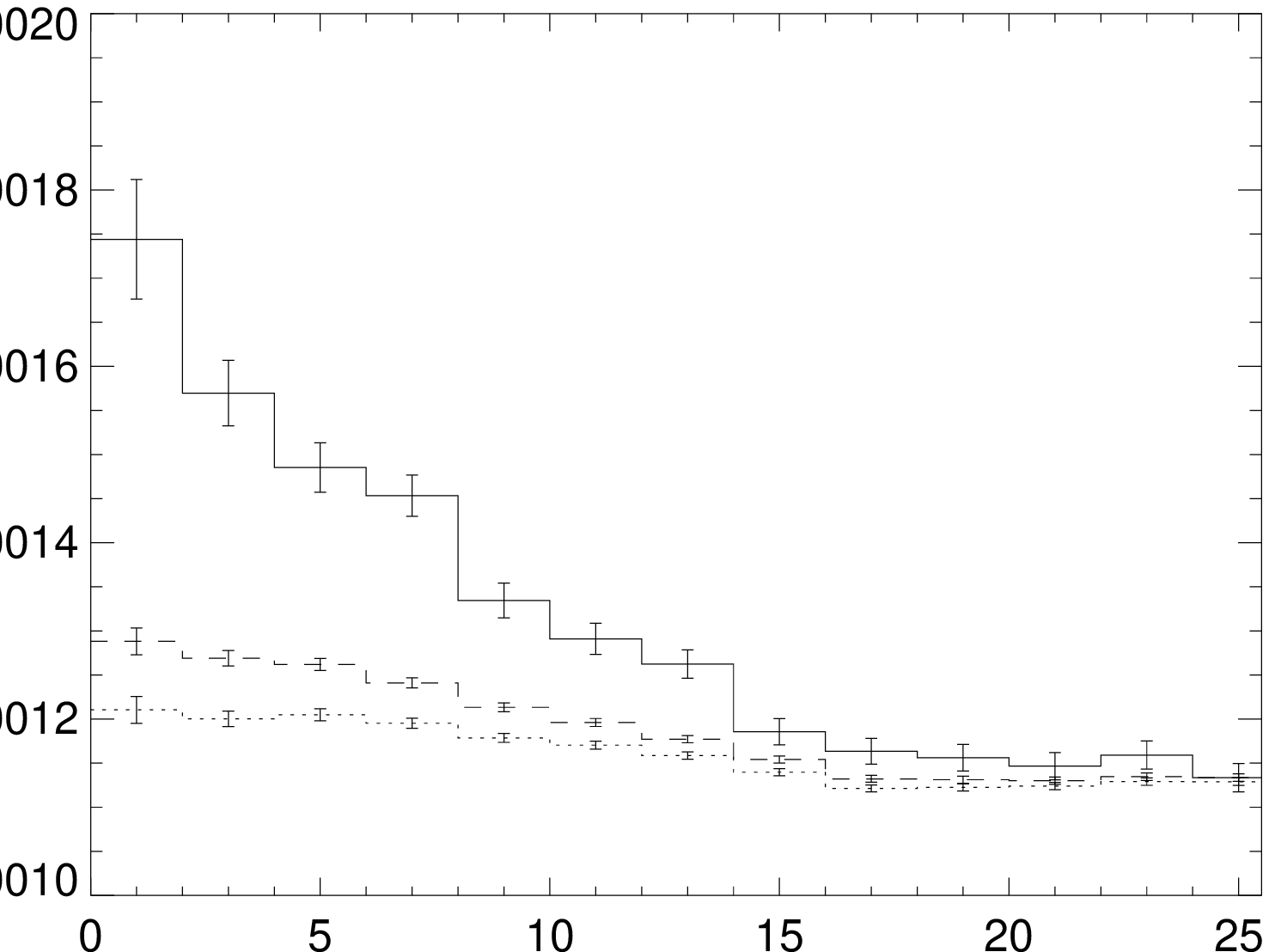}{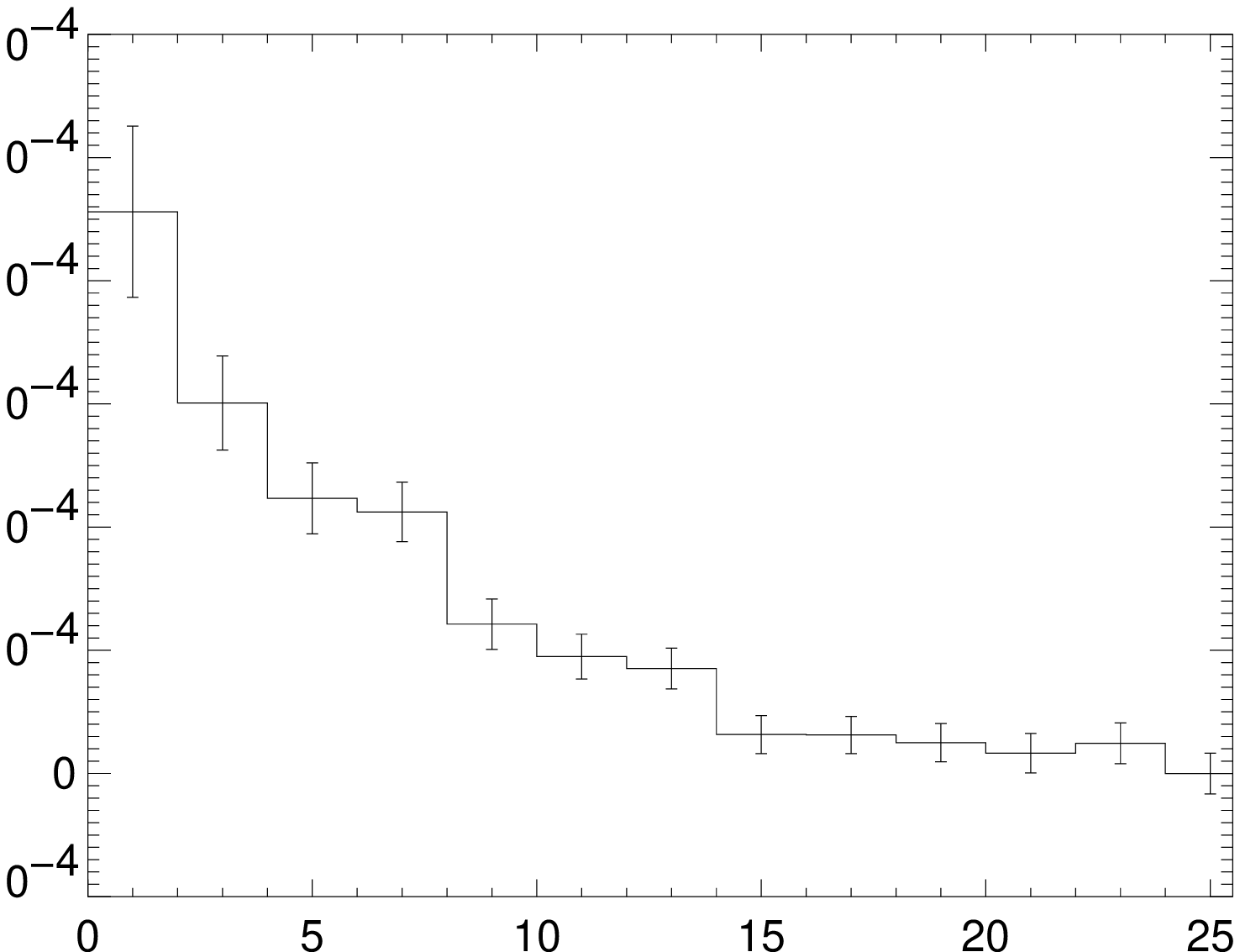}
\caption{
\captionone
\label{radial}}
\epsscale{1.0}
\end{center}
\end{figure}
}

The identification of diffuse emission in a 
sky map is difficult because of the low 
signal to noise ratio of data in individual cells of 
the map. This problem was identified early 
in the study of diffuse X-ray emission in 
clusters of galaxies. A solution univerally employed
in studies of diffuse emission in 
clusters is to construct the azimuthally averaged 
radial intensity profile of the flux. We 
derived this profile for the EUV emission in 
the Coma Cluster. The results are shown in 
Fig. 1.  There is a substantial EUV excess 
out to  $\sim 14\arcmin$
and marginal evidence for
emission to $20\arcmin$.  The dominant uncertainty in the determination of the
overall excess is uncertainty in the determination of the ratio between the
EUVE and ROSAT count rates due to the X-ray plasma.
Including this uncertainty, the overall EUV excess is significant at greater 
than
the 12$\sigma$ level.

In order to obtain a value for the total EUV excess in physical rather
than instrumental units, we summed the excess counts shown in Figure 
\ref{radial} and
computed the unabsorbed count rate by correcting for Galactic 
interstellar absorption
as described above. We then divided by the EUV instrument effective 
area to
obtain results in physical units.  The unabsorbed EUV 
excess in the band from 68 to 92
\AA\ (the approximate bandpass of the observation as
defined by the telescope high energy cutoff and the
low energy cutoff produced by Galactic absorption) is
$1.7\times 10^{-13}$  \unit{erg s^{-1} cm^{-2}
\AA^{-1}}. Assuming a distance of 100 Mpc, this
corresponds to a total energy output between 68 and 92
\AA\ of $4.9\times 10^{42} {\rm erg s^{-1}}$. In this
calculation we assumed spectral indices between 1 and
1.6 
which are appropriate given the source mechanism
identified for the emission as discussed below. This
result is relatively insensitive to the spectral
index, with a variation in the flux of only a few
percent for the index range listed.
It is of interest to compare this energy output with
the energy output of the X-ray plasma which is about $10^{45}$ 
ergs/s, based upon a central density of $3\times 10^{-3} \percmcubed$,
a core radius of 10\arcmin.5, a $\beta$ of 0.75, a temperature of 9 
keV,
and the cooling function of
Sutherland and Dopita 1993.
If the EUV excess were due to a thermal plasma at 10$^6$ K, the 
bolometric luminosity of this plasma would be $5\times 10^{44}$ 
erg/s,  which is
comparable to the energy output of the X-ray plasma. 

The value we obtain for the EUV excess, $F_\lambda$, is about a factor of 
two smaller 
than the number reported by Sarazin and Lieu 1998 after correcting for 
a
difference in the assumed distance to the Coma Cluster.
Although
it is impossible to conclusively identify the reason for this 
difference,
we note that if we were to inappropriately compute the energy output using the full bandpass of
this instrument rather than the effective bandpass, we would obtain
a value similar to that reported by Sarazin and Lieu.

\section{The Relationship between the EUV Excess and the X-ray 
Emission}

\ifpp{
\begin{figure}[tb]
\begin{center}
\plotone{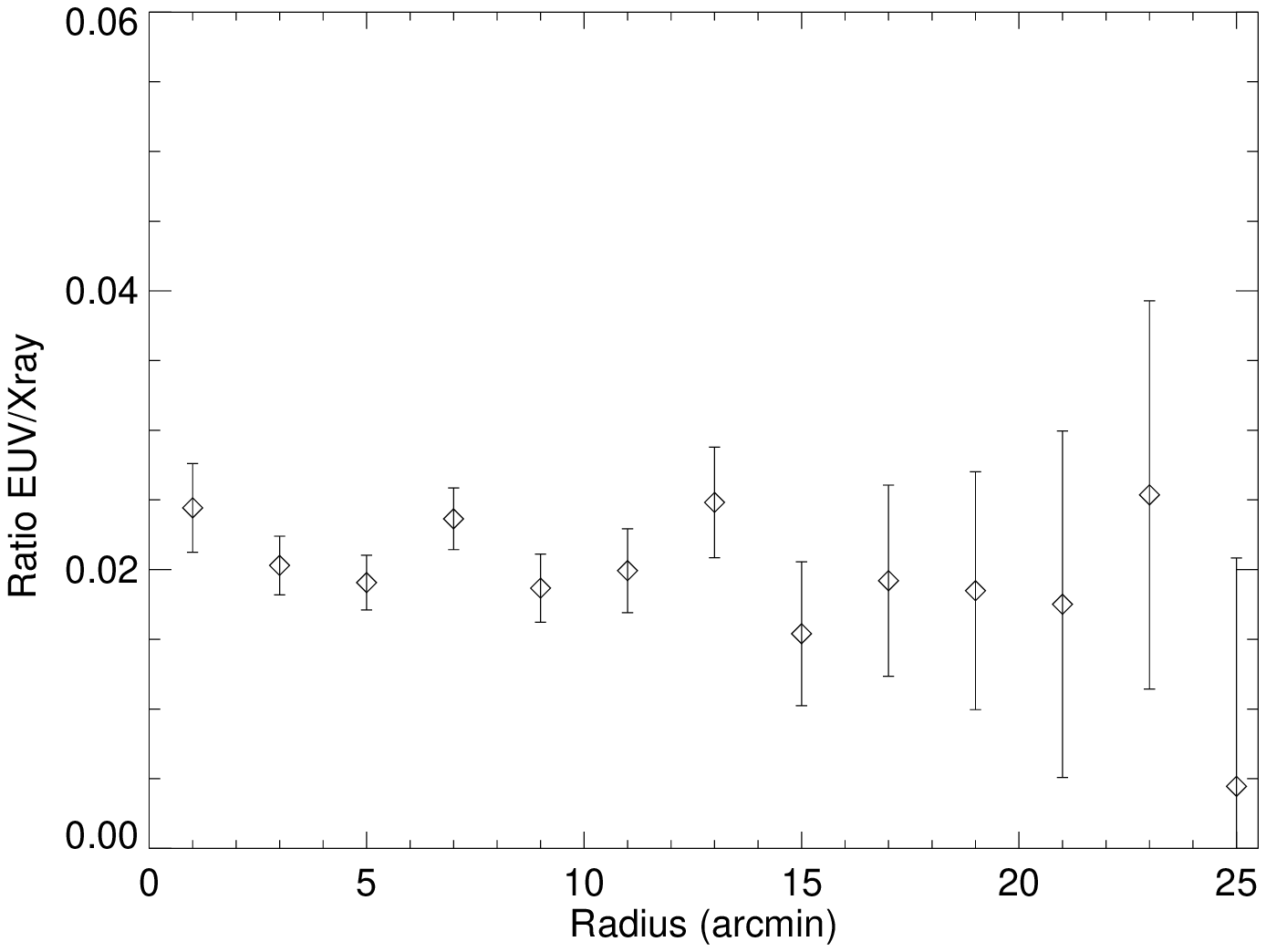}
\caption{
\captiontwo
\label{ratio}}
\end{center}
\end{figure}
} 

We first derived the ratio between the azimuthally averaged EUV flux 
(0.13 - 0.18 
keV) and the azimuthally averaged X-ray flux derived 
from the ROSAT 0.5-2.4 keV X-ray band. We show this ratio as a 
function 
of increasing distance from the 
cluster center
in Figure \ref{ratio}.   As can be seen this ratio is essentially flat.  
The error bars
increase at larger radii because of the limited EUV excess flux at 
these larger
radii.

An azimuthally averaged radial intensity profile is quite sensitive to 
the presence of diffuse emission. However, by its very 
nature this process eliminates any possibility 
of examining details of the spatial 
distribution of the emission other than its average radial 
distribution. 
A direct study of the details of the EUV 
emission in this bandpass can only be achieved with a very substantial data set, which is now unobtainable.  
As an alternative, we considered 
ways to investigate aspects of the spatial distribution which might 
prove to be useful. We first considered 
the number of EUV excess counts in individual cells in the 
sky map. The telemetered cell size of EUVE 
data is $4\arcsec .6$.  We summed these data 
into larger blocks. The minimum 
appropriate cell size is $0\arcmin .28$  because the 
registration of the EUV images are 
uncertain at this level. In addition, the 
use of a cell size smaller than the intrinsic 
resolution of the telescope could potentially 
provide misleading results. The response of 
the telescope is closely replicated by a 
Gaussian with a 90\% included energy width 
of 1\arcmin\ and one possibility
would be to convolve the data with a
Gaussian of this size. However, we summed the counts in a 1 min square box 
since Hardcastle 2000 has pointed out that the use 
of a smoothing function adds considerable uncertainty to the  
significance levels of the resultant data set. 
We then computed isophotes of the EUV 
excess. The results are shown in Figure \ref{map}.     
The EUV excess appears to be more extended to the 
southeast 
although this is a
region of low counts per bin.

\ifpp{
\begin{figure}[tb]
\begin{center}
\plotone{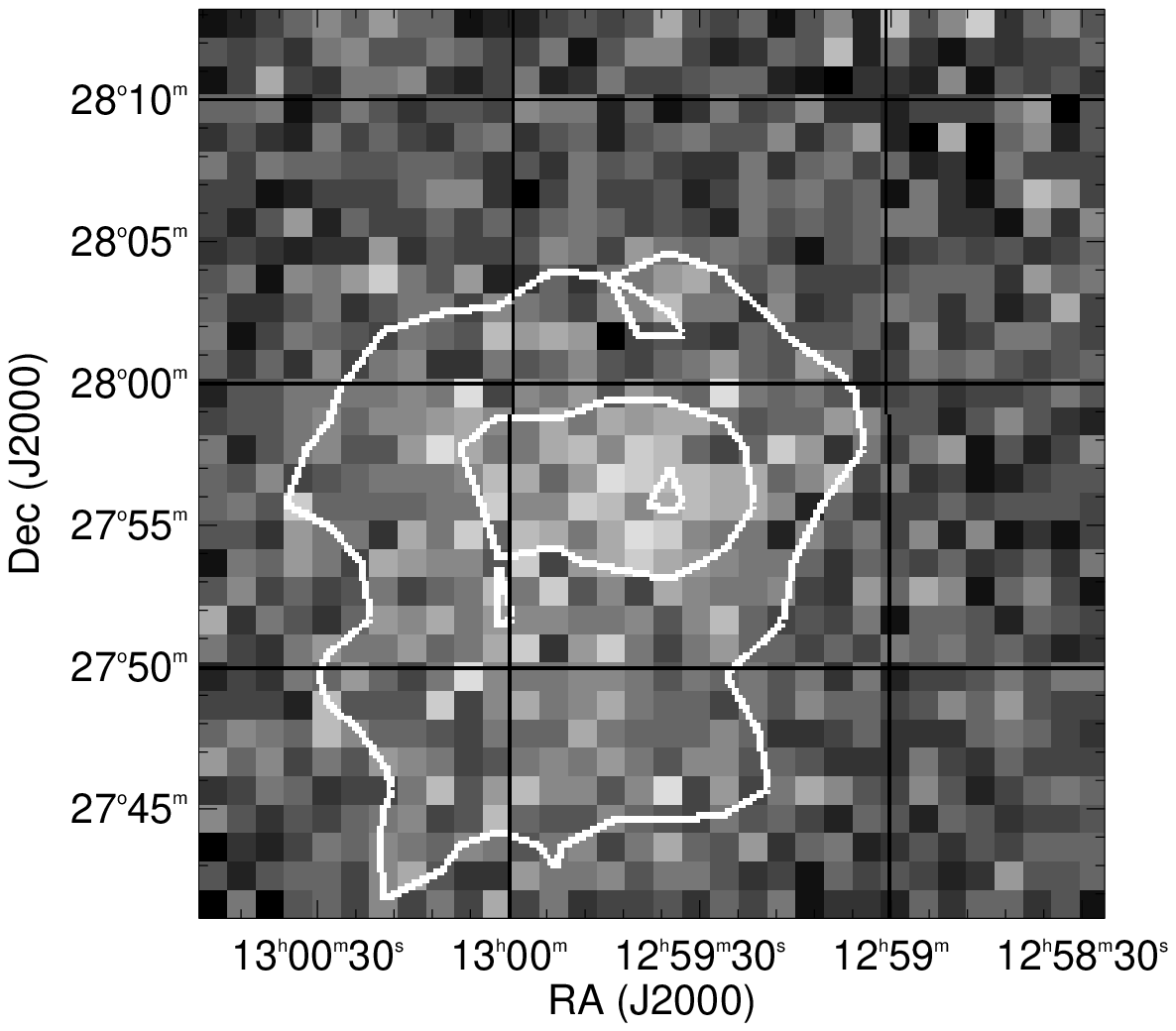}
\caption{
\captionthree
\label{map}}
\end{center}
\end{figure}
\begin{figure}[tbp]
\begin{center}
\plotone{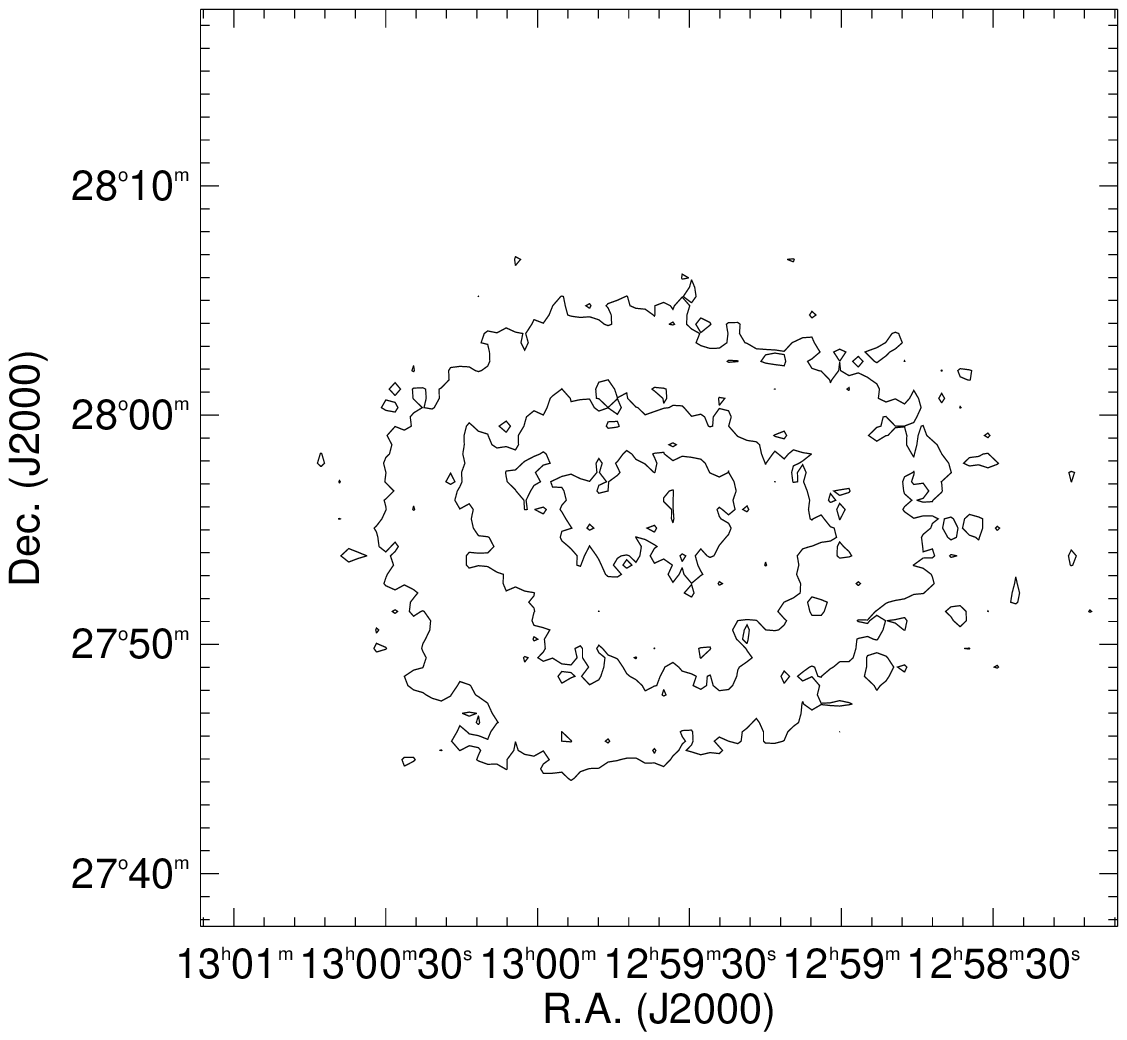}
\caption{
\captionfour
\label{xray}
}
\end{center}
\end{figure}
}

In Figure \ref{xray} we show the X-ray 
emission in the Coma Cluster
derived from archival ROSAT PSPC
data. 
A comparison of Figure \ref{xray} with Figure \ref{map} shows that the 
EUV emission is only
detectable in the central portion of the X-ray image.  This may be 
because the
EUV emission is only present in the core, but it could 
simply be
a sensitivity issue, and the lower intensity wings extend further out.   
We then carried 
out a
standard linear correlation analysis between
the EUV excess dataset shown in Figure \ref{map} and the corresponding 
X-ray data shown
in Figure \ref{xray}.

An immediate problem in carrying out a correlation 
analysis between the EUVE excess and the 
X-ray emission is that 
a correlation analysis will compare the 
number of counts in a given cell in one 
image with the number of counts in an 
identical cell in the other image without 
accounting for any statistical fluctuations in 
these values. Hence with an image with a small number of counts in 
individual cells, a false statement of a lack of correlation will be 
provided simply because of the statistical fluctuations of the data in 
the cells.  

To assess this effect for the EUV excess, we carried out a correlation 
between 
two independent data sets of the EUV 
excess in the Coma Cluster 
as we summed the counts in 
individual cells into larger sized bins. The 
results are shown in Fig. \ref{correlation} as 
diamonds. 
As expected, the correlation between the
two independent data sets of the EUV 
excess is quite poor with smaller bin 
sizes, but increases rapidly as the bin sizes are increased and more 
counts are registered in each bin.

We note that the determination of the
confidence levels of the correlation measures shown in Fig. \ref{correlation}
are inherently
complicated by three statistical properties of correlation estimates: 
(a) they are
inherently non-Gaussian, being mathematically bounded to the interval
-1 to +1;  (b) they are asymmetrical;  (c) their confidence intervals 
depend
on the true population correlation value, which is unknown.   This 
situation
prompted Fisher 1935
to create a nonlinear transformation of the correlation statistic into 
a 
Gaussian
normal variate with uniform variance, namely
{\em Fisher's transformation}
which depends only on the observed correlation value $r$ and 
the
number of independent data points $N$, both of which are known.  In 
use,
one simply converts the observed $r$ into Fisher's $z$ (which is 
Gaussian) 
and
then converts the desired confidence interval in $z$ back into an 
interval 
for $r$.
Using this method we obtain the error 
bars shown in 
Figure \ref{correlation}.  We confirmed these error values by 
performing multiple Monte Carlo
simulations of uncorrelated data.  The standard
deviation of the correlation coefficients of these uncorrelated 
simulations
in Fisher $z$-space was equivalent to
the error values calculated using the number of independent points.

We expected that there were 
sufficient counts in the deep ROSAT X-ray 
image that uncertainties in the photon 
statistics in the X-ray data would be 
inconsequential in comparison with the uncertainties due to the limited 
data in the EUVE data set. A self-correlation of the X-ray data verified 
this conclusion. 

\ifpp{
\begin{figure}[tb]
\begin{center}
\plotone{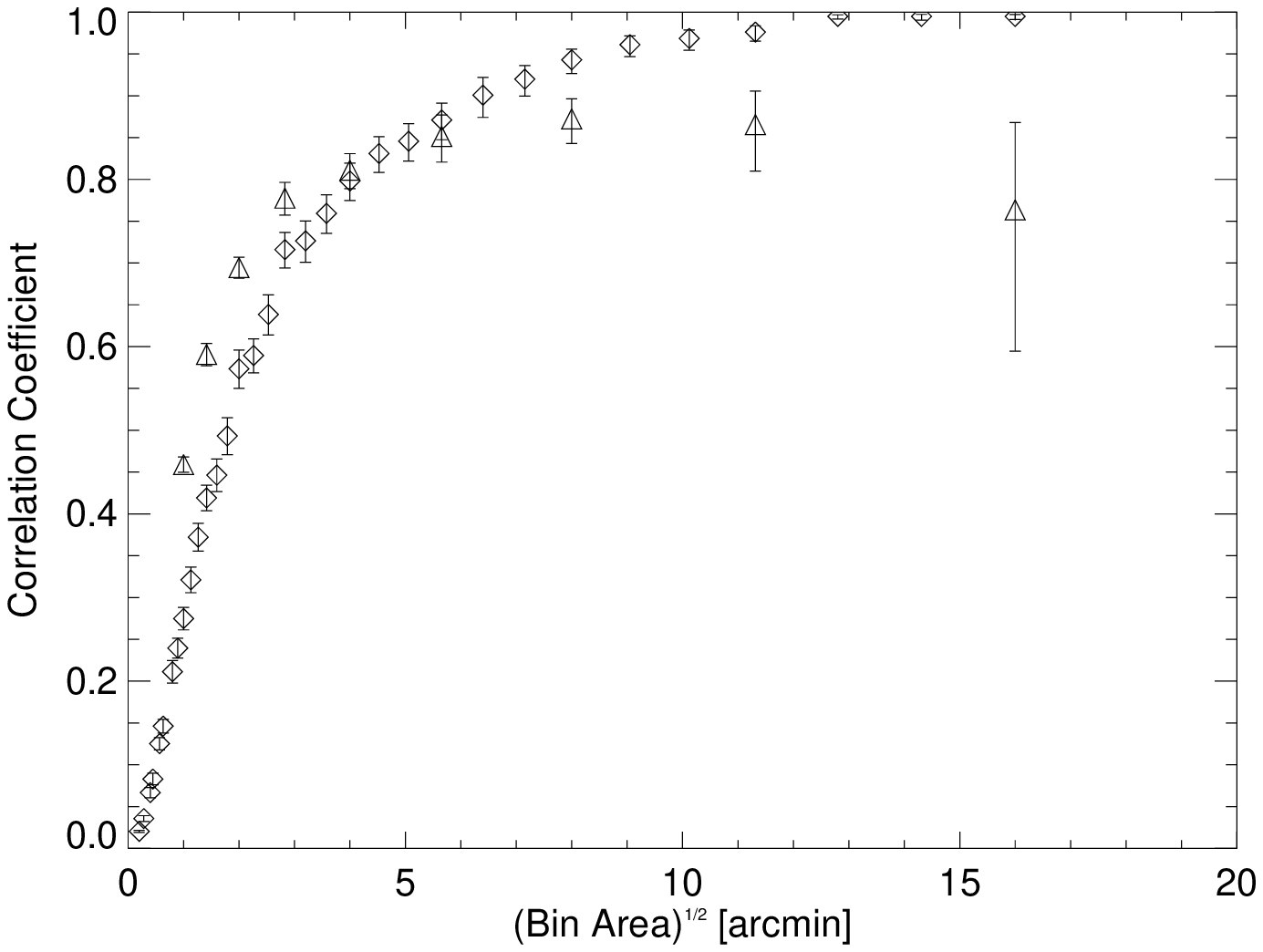}
\caption{
\captionfive
\label{correlation}
}
\end{center}
\end{figure}
}

We then carried out a correlation of the EUV excess 
image with the X-ray image  
as a function of 
increasing cell size.  The
results are shown in Figure \ref{correlation} as triangles. Up to a 
scale of 
4.0 arcmin$^{2}$ the EUV self correlation and the 
EUV/ X-ray correlation both rise 
reflecting the limited quantity of EUV data. 
At larger scales the EUV self correlation is 
better than the EUV/X-ray correlation 
allowing us to make several definitive statements. 
First, there is a substantial, but not exact, spatial similarity 
between the EUV and the X-ray 
emission. This similarity peaks at scales of 
$8.8$ arcmin$^2$ at a value of 0.86. There is an indication that the 
correlation 
falls off at 
larger scales, though this decrease is not significant at the three 
sigma level. 
We can also conclude that at scale 
sizes smaller than $4.0$ arcmin$^2$ the correlation is no better than 
0.80 since 
a correlation cannot be improved by simply reducing larger image pixels 
to smaller pixels.
 
In summary, we can state that at scales greater than $4.0$ arcmin$^2$ the 
spatial 
distributions of the EUV and the X-ray emission have substantial, but 
not exact, similarities. The correlation peaks at a value of 0.86 at a 
scale size of $8.8$ arcmin$^2$ with a suggestion that the correlation 
decreases 
at a larger scale. At scales less than $4.0$ arcmin$^2$ the correlation 
is no 
better than 0.80.

\section{Discussion}

In view of these new findings, it is useful to 
reconsider suggestions for the 
underlying source mechanism for the EUV 
excess in clusters of galaxies. One proposal 
was that this emission was thermal emission 
from a ``warm'' (10$^6$ K) gas (Lieu \etal 
1996a; 1996b; 1999a,b; Mittaz \etal 1998; 
Bonamente \etal 2001). Buote 2000a,b carried out extensive analyses 
of ROSAT PSPC data and also claimed to have found evidence for gas at this 
temperature in the core of several clusters. 

The EUV emission in the Coma Cluster is 
almost spherical and could conceivably be the 
product of a gravitationally bound gas. However, the maintenance of a 
warm intracluster gas 
is quite difficult to understand since gas at 
this temperature is at the peak of its cooling 
curve and would typically cool in less than 0.5 Gyr  (Landini \& 
Monsignori Fossi 
1990). 
This has resulted in a fair level of skepticism in regard to a thermal 
origin for the EUV excess. A variety of observational studies have been 
carried out in an attempt
to discover evidence for a warm $10^{6}$ K thermal gas.   Initial 
studies
with XMM-{\em Newton} showed no lines from a
$10^6$ K gas in any of the clusters examined (Peterson \etal 
2001). 
However the 
Coma Cluster was not examined in these studies, which raised at least 
the possibility that Coma is uniquely different and that the EUV 
excess in this cluster is indeed thermal in origin. Dixon 
\etal 2001 obtained long Far Ultraviolet Spectroscopic Explorer (FUSE) observations centered on the Coma 
cluster in search of O{\sc vi} 1032, 1038 emission which would be 
produced 
by  a 10$^6$K thermal gas. Although this emission was detected, the 
high 
resolution of FUSE showed that all of this emission was Galactic and that
none was red-shifted O{\sc vi} from material in the cluster. 
However, the presence of thermal gas could not be completely ruled out
since a large depletion of oxygen in the 
cluster
would render these lines unobservable. 

Additional information on this topic has been obtained
by Arnaud et al. 2001 and by Vikhlinin et al. 2001.
Both of these groups studied the core of the Coma
cluster in detail. Arnaud et al.~analyzed XMM-{\em Newton} data and
determined temperatures in 3.5$\arcmin \times 3.5\arcmin$ regions 
in the $20\arcmin$ 
core of the cluster. The temperatures in these regions
ranged from 7 to 8.5 keV (with a few outliers) with no
evidence for lower temperature gas. Vikhlinin et al.
used {\em Chandra} observations to search for lower
temperature gas in very small regions in the core of
the cluster. They found 1 to 2 keV gas within a 7$\arcsec$
radius of NGC 4874 and NGC 4889 which they attributed
to emission from the halos of these individual galaxies. However,
immediately outside of these small regions they found
only high temperature (9 keV) gas. 

Finoguenov et al. 2003 used XMM-{\em Newton} data in $\sim 20\arcmin$  
diameter bins
and detected O{\sc vii} and O{\sc viii} emission $\sim 30\arcmin$
off-center from the core of the cluster. These
lines would be produced by a 0.2 keV (or $2 \times 10^6$ K)
gas. They showed that this emission came from a filament in
front of the Coma Cluster which was seen in projection
against the cluster. The key difference between these
measurements and the measurements of Vikhlinin et al. and Arnaud et al. 
was the fields of view involved. Only with the larger field of 
view is the line
emission from the low temperature gas detected.
Finoguenov et al. found that the oxygen line emission was 1/30th of
the X-ray emission of the 9 keV gas in the cluster center. 
Even if Finoguenov et al. were incorrect in their
interpretation that this emission was from a filament in front of
the cluster and was, instead, associated with 
$2 \times 10^6$ K gas in the core of the cluster, its EUV
emission would have been so faint as to be
unobservable with EUVE and could not have been 
responsible for the excess reported here. 

Kaastra, Lieu et al. 2003 claimed to have found
``warm'' thermal emission 
within the central 12$\arcmin$ of the Coma Cluster with XMM-{\em Newton}.
However, their claim is based on the (marginal) detection of a diffuse
soft X-ray excess in the cluster and not on the detection of oxygen lines
and it is contradicted by the work of Arnaud et al. 2001 and Vihklinin et al. 2001.
The Kaastra, Lieu et al. 2003 results could equally well be interpreted as non-thermal emission.

Since the EUV flux is not thermal in origin, we have examined non-thermal processes as the source of this emission.
Inverse Compton (IC) scattering 
of the 2.7 K cosmic microwave background (CMB) photons by energetic 
electrons 
(Hwang 1997; Bowyer \& Bergh\"ofer 1998; En\ss lin \& Biermann 1998; 
Sarazin \& 
Lieu 1998) was suggested early on, and it is still
the only suitable candidate non-thermal mechanism (Blasi \& 
Colafrancesco
1999; Atoyan \& V\"olk 2000; Brunetti et al. 2001a; Petrosian 2001; 
Tsay \etal
2002). 

En\ss lin, Lieu \& Biermann 1999 suggested cluster starlight radiation
as the background photon field.
The energy density in starlight photons is about two orders
of magnitude less than that in the CMB in the core of Coma, so the
starlight-IC model requires a number density of $\sim 5$ MeV electrons 
which is 
comparable to the thermal electron density (En\ss lin et al. 1999). These 
particles 
then provide the dominant pressure in the cluster. This condition seems 
implausible  
both to establish and to maintain. If such
low energy electrons were mixed with the thermal plasma, they would
transfer their energy on timescales of a few hundred million years by
way of Coulomb collisions which would result in excessive heating, even
ignoring heating by likely associated non-thermal protons.
Magnetic fields strong enough to separate these non-thermal particles
from the thermal plasma would lead to magnetic pressures exceeding
the thermal gas pressure, which is similarly unlikely. If the posited 5 
MeV electrons were relics of a much more energetic population, their
original energy content would have been much greater than
that already required for the starlight-IC model itself, making these 
problems worse. Thus, we reject this hypothesis as an untenable 
explanation for the EUV excess in Coma.

A number of authors have suggested
specific IC-CMB models for the production of the EUV excess.
Atoyan \& V\"olk 2000 posited a population of ``relic''
electrons driven into the intracluster medium by galactic winds during intervals of
galactic starbursts and then reaccelerated by strong merger shocks. To
avoid excessive radio emissions from the assumed hard energy spectrum
of the EUV electrons ($s_e = 2.1$)
with multi-$\mu$Gauss magnetic fields, their model included an {\em ad 
hoc}
electron energy cutoff near 250 MeV. Brunetti et al. 2001b proposed
a model for EUV IC-CMB emissions based on turbulent reacceleration
of non-thermal
electrons recently injected by the head-tail radio galaxy NGC 4869, 
which is
several arcminutes west of the cluster center.
This was an extension of a model that would
explain the radio emission and the 40 KeV X-ray 
emission in the Coma Cluster (Brunetti \etal 2001a). 
Important constraints imposed were that 
this population should not produce (an unobserved) $\gamma$-ray 
emission, 
and it should produce the observed spectral steepening of the radio 
emission with increasing distance from the cluster core. This model 
included an initial cosmic ray population produced $\sim~$2Gyr ago, 
reacceleration $\sim~$1Gyr ago by shocks from mergers, and a recent 
injection of low energy cosmic rays that is responsible
for the EUV emission. 
Their model assumed a relatively
hard electron spectrum for the EUV electrons (hereafter, EUVe) with $s_e = 2.6$, and a magnetic 
field 
$\sim 0.5$ $\mu$Gauss, and required a cutoff near 500 MeV
to avoid emissions conflicting with observations in other bands.
The relative complexity of the full model
illustrates the difficulty in finding a unified
model for non-thermal emissions in Coma.  

Sarazin \& Lieu 1998
proposed a model for EUV emission in clusters
in which relic low energy electrons accumulated from various
origins would be distributed similarly to the thermal plasma,
that is $n_{EUVe} \propto n_{te}$. That distribution
predicted an azimuthally averaged ratio, $I_{EUV}/I_X \propto 
1/n_{te}$,
increasing substantially with distance outside the cluster core. 
Bergh\"ofer
\etal 2000 derived this ratio for the Virgo cluster as a test of this 
model.
They found that the ratio was flat with increasing distance from the 
center of the
cluster in contradiction to the prediction of this model.
We have derived
this ratio for the Coma Cluster, and, as shown in Figure \ref{ratio}, 
it is
essentially constant. Using a beta model for Coma (Briel et al. 1992)
with $\beta = 0.75$, $r_c = 10\arcmin.5$, and assuming
$n_{EUVe} \propto n_{te}$ one finds that $I_{EUV}/I_X$ should
have increased by about a factor of two 
from the cluster center to a distance of 10\arcmin.5 and by a factor of 
six at
21\arcmin. All of these outcomes
are clearly inconsistent with the data. 

Three previous studies considered secondary emission
in connection with the EUV excess. Blasi \&
Colafrancesco 1999 considered secondary emission as
part
of a unified model for non-thermal emissions in
Coma. They found their model had multiple problems.
The spatial distribution of the radio emission was not
correct, and too much gamma radiation was produced.
Finally, the EUV emission produced was too low. Blasi
2001 modeled emission from secondaries as part
of a treatment of nonthermal emission in cluster
mergers, including Coma. He assumed that a strong
merger shock would inject primary electrons and
protons with a density distribution proportional to
the thermal plasma. With the parameters he employed,
the associated EUV emission would be dominated by
primary electrons, so the resultant EUV spatial
distribution would take the same form as that proposed
by Sarazin and Lieu (1998), and would be incompatible
with the observational results reported here. Miniati
et al. 2001b estimated the EUV flux from secondary
emission in clusters as part of a larger study of
cluster formation. Their 
EUV luminosity vs cluster temperature relations
underestimated the observed EUV flux in Coma by about
an order of magnitude. Consequently, Miniati et al.
did not pursue the idea that IC-CMB emission from
secondary electrons could be the underlying source
mechanism for the EUV excess. In retrospect, their
luminosity estimates were artificially low because of the
effects of finite numerical resolution in their
simulations.  This significantly reduced the central
gas densities in clusters which resulted in an
underestimate of the secondary emission flux.

No existing models produce an EUV intensity distribution that is highly
correlated with the thermal X-rays as shown in Figure 
\ref{correlation},
and simultaneously produce a constant ratio
between the azimuthally averaged EUV and X-ray intensities
as illustrated in Figure \ref{ratio}. Therefore,
we have searched for a new model that would naturally produce these outcomes.

\subsection{General Constraints}

Before introducing a specific model that will yield these observational findings, we first establish some general constraints on the emitting particles and their environment that would apply to any successful model. In the IC-CMB scenario, the EUV excess
is produced by electrons of characteristic
energy $E \sim 200\sqrt{\epsilon_{150eV}}$ MeV 
($\gamma \sim 400\sqrt{\epsilon_{150eV}}$), where $\epsilon_{150eV}$ 
is the EUV photon energy, normalized to 150eV ($\lambda \sim 80 \AA$).

The magnitude and distribution of the magnetic field in the cluster are
important constraints in any model for non-thermal emissions in Coma.
Extensive work has been carried out in efforts to determine the magnetic
field strength.  Recent summaries of the observational situation and possibility
of reconciling the (apparently) contradictory results have been provided by
Kronberg 2003; Clarke 2003; and Brunetti 2003.
Different approaches yield different results. One approach 
is to calculate the field based on the assumption of equipartition 
between the energy density of the relativistic particles associated with
the radio 
emission and the magnetic field. 
A recent and especially detailed 
result for the Coma Cluster using this approach has been 
obtained by Thierbach et al. 2003 who 
find an equipartition field of $\sim 0.7 \mu$Gauss  if electrons are the relativistic gas, 
or $\sim 1.9 \mu$Gauss if the proton-to- electron energy density
ratio in the relativistic gas is the same as that in the 
ISM.
Faraday rotation measures of radio sources in clusters have been 
extensively studied as a means of determining cluster magnetic fields. 
Very high fields have been obtained using sources embedded 
in the cores of clusters with cooling centers (Eilek 1999; Taylor et 
al. 1999). Rotation measures of radio sources behind clusters have been 
measured by a number of groups (Kim et al.~1990; Feretti et al.~1995). The most extensive results using this approach have been obtained by 
Clarke et al.~2001. They find fields that are typically in the range of 
5 to 10 $\mu$Gauss. 

Fields in the range of 0.1 to 5 $\mu$Gauss are required if
the EUV excess is the product of IC-CMB in a uniform
magnetic field (e.g. Hwang 1997; Atoyan \& V\"olk 2000; 
Brunetti et al. 2001b; our discussion below), but the
higher values in this range can only be realized if a
rather arbitrary high energy cutoff is imposed upon
the underlying cosmic ray spectrum, or if the cosmic
ray spectrum is very steep. A less extreme explanation
for the lower fields required in IC-CMB models for the
EUV is that the magnetic fields are not homogeneous
and that the EUV excess originates in low-field regions
while high-field regions produce the higher Faraday
rotation measures (Petrosian 2001; Newman et al. 2002;
Beck et al. 2002). Tregillis et al. 2003 studied
synthetic radio and X-ray images derived from high
resolution three-dimensional MHD radio galaxy simulations to compare
average field estimates with actual magnetic field
properties in the simulated objects.  They found that the estimated fields
roughly corresponded to actual rms fields, but scattered around the physical rms
value by a factor of $\sim$2-3.  In light of the above discussion, we assume a
field of about 1 $\mu$Gauss in the ICM of the Coma Cluster in the following.

Energy loss timescales provide an important general constraint on 
models for the EUVe. For conditions in the X-ray core of Coma, approximately coinciding with the EUV excess
(thermal electron density, $n_{te} \approx 3 \times 10^{-3} {\rm 
cm}^{-3}$, Briel et al. 1992),
IC-CMB and Coulomb energy losses are currently roughly comparable at 
200 MeV
(e.g. Sarazin 1999; Petrosian 2001). However, since synchrotron energy 
losses
compare to IC losses by
the ratio $(B/B_{\mu})^2$ and  $B_{\mu} \propto (1+z)^2$, this ratio 
was probably smaller
in the past. With a magnetic field of about 1 $\mu$Gauss,
synchrotron losses can be neglected.
Using standard relations (e.g. Sarazin 1999; Petrosian 2001)
and correcting the IC-CMB loss rate for the Hubble expansion with
$q_0 = 0.5$, but assuming $n_{te}$ was not greater in the past,
it is simple to demonstrate that IC-CMB losses were
dominant over Coulomb losses at these energies, and 
that
the characteristic IC-CMB energy-loss lifetime for EUVe
is $\sim 2$ Gyr.

Another key point is that electrons at these energies diffuse very slowly 
in the intracluster medium 
(Schlickeiser \etal 1987; V\"olk \etal 1996) For
Bohm diffusion in a $\mu$Gauss field, 200 MeV electrons would diffuse
only about 10 pc during their lifetimes. In almost any plausible
cluster field and turbulence model the EUVe 
are effectively tied to the local plasma.
Mixing of the cluster plasma will take place on
timescales of Gigayears in response to stirring in the
cluster (e.g. Markevitch, Vikhlinin \& Mazzotta 2001)
caused by mergers and AGN activity. But since the EUVe
electrons are tied to the cluster medium, the spatial
distribution of both of these species will be similar,
though not identical. In clusters with current active
energy deposition such as Virgo or Hydra, fresh
particle populations will not become mixed
immediately, as illustrated by the X-ray holes seen in
such clusters (e.g. Nulsen \etal\ 2002). On Gigayear
timescales, however, the nonthermal particles will
become mixed throughout the cluster.

We next explore spectral constraints on the
EUVe population that can be derived by requiring that it
does not produce emission in other bands that exceeds
those observed. The EUVe population directly includes
only energies near 200 MeV, but is likely to continue
to higher energies following a normal power-law
spectrum. In particular, we see no reason to introduce
an artificial cutoff at higher energies. Clearly an
important constraint is that the high energy extension
of the EUVe population does not produce IC-CMB in
excess of the observed non-thermal high energy excess
(hereafter HRX). Both BeppoSax (Fusco-Femiano et al.
1999) and RXTE (Rephaeli \& Gruber 2002) gave results
that can be expressed in terms of a flux near
40 keV of $\nu F_{\nu} 
\approx 8\times 10^{-12} {\rm~erg~cm^{-2}s^{-1}}$ inside a radius
$\sim 1$ degree. If the EUVe
population includes electron energies approaching 4 GeV, its 
IC-CMB spectrum will reach
into this band. Our measured EUV flux corresponds to
$\nu F_{\nu} = 1.4 \times 10^{-11} {\rm~erg~cm^{-2}s^{-1}}$
at 150 eV. A simple power-law extension of the
IC-CMB spectrum from 150 eV to 40 keV with a
spectral index, $\alpha$, ($F_{\nu} \propto \nu^{-\alpha}$,
corresponding to an electron energy distribution
$n_e(E_e) \propto E_e^{-s_e}$, where $s_e = 2\alpha + 1$)
would fall below the observed excess 40 keV flux if $\alpha > 1.1$,
or $s_e > 3.2$. Since the HRX field is substantially larger than the
EUV source, the extended spectrum could be steeper,
so $s_e > 3.2$ is a conservative limit above $E_e \sim 200$ MeV.  

Similarly, extension of the EUVe population to
higher energies
could contribute detectable radio emission. This 
constrains
both the form of the electron spectrum and the effective magnetic field 
strength, as discussed previously. The lowest frequency synchrotron flux measured
for Coma C is
49 Jy at 30.9 MHz (Giovannini \etal\ 1993). The radiating electrons
would have characteristic energies $E_e \approx$ 1.4 GeV 
$B^{-1/2}_{\perp}$, where $B_{\perp}=B(\cos\theta)$ is the sky-plane 
component of the source magnetic field expressed in $\mu$Gauss.
For $B_{\perp} \sim 1$ $\mu$Gauss, these electrons
would be roughly
an order of magnitude more energetic than those
producing the EUV
emission. The requirement that the
observed radio flux exceed any synchrotron flux,
$F_{\nu_s}$, produced 
by a high
energy extension of the EUVe population can be
conveniently
expressed by the constraint
$R_{si} \equiv \nu_s F_{\nu_s}/\nu_i F_{\nu_i}
=\lambda_s
F_{\lambda_s}/\lambda_i F_{\lambda_i} < 1.07\times
10^{-3}$,
where $F_{\nu_i}$ is the observed IC-CMB flux in the
EUV.

Assuming a power-law electron energy
distribution
over the relevant range, and that the EUV emission
is IC-CMB, the ratio of
the associated radio synchrotron flux to the EUV
flux in a uniform magnetic field is easily shown to be
(Jones \etal\ 1974)
\begin{equation}
B_{\perp} =  \left[
\frac{\pi}{3} \frac{j^{bc}_{\alpha o}}{j_{\alpha o}}
\left( \frac{\nu_{\mu} \nu_s}{\nu_i \nu_{\mu
B}}\right)^{\alpha - 1}
R_{si}\right]^{1/(1+\alpha)} B_{\mu},
\label{synch-ic}
\end{equation}
where $\nu_s = 30.9$ MHz and $\nu_i = 37$ PHz are
the observed 
frequencies for
synchrotron and IC-CMB emission, $j_{\alpha o}\sim
1$ and $j^{bc}_{\alpha o}\sim 1$
are constants tabulated in Jones \etal\ 1974,
$\nu_{\mu} = kT_{CMB}/h = 
57~(1+z)~$GHz,
$B_{\mu} = 3.2~(1+z)^2~$ $\mu$Gauss is a fiducial
field strength
whose energy density matches the CMB, and $\nu_{\mu B} = eB_{\mu}/(2\pi 
mc) = 9~ (1+z)^2$ Hz.

A range of values for the integrated Coma C radio spectrum is viable; 
the appropriate choice depends on the (unknown) underlying source 
model (Thierbach et al. 2003). Reasonable values for the  spectral index 
range from $\alpha = 1.35$
to 0.83. For consistency, values of $R_{si}$ 
near the 
observed limit would require model spectra at least as steep.
The associated magnetic field constraint from equation \ref{synch-ic}
depends on the spectrum chosen. For example, it would range from $B_{\perp} \sim 0.1~\mu$Gauss for 
$\alpha 
= 1.0$ ($s_e = 3$) to $B_{\perp} \sim 1.4~\mu$Gauss for $\alpha = 1.5$ 
($s_e = 4$). 
If the magnetic field is isotropically
oriented, these values  of $B_{\perp}$ should be increased by roughly 
20\% to
arrive at an estimate of the magnetic field.  

\subsection{Secondary Electrons as the Source of the EUV Flux in Coma}

We now propose a specific model that produces the observational
results presented here, and then demonstrate that this model does not
violate the general constraints derived above.
The observed correspondences between the EUV and
thermal X-rays strongly 
suggest that the low energy EUV emitting particles
and the thermal 
intracluster medium have a physical interdependence.
As pointed out by Sarazin and Lieu 1998, the intuitively obvious 
relation between cosmic rays and thermal plasma is one 
in 
which the cosmic ray particles are relatively well mixed with the 
thermal plasma, \ie\ in which $n_c \propto n_t$.
This would be a natural expectation if the
cosmic rays were 
accumulated over much of the formation of the
cluster, and its various constituent components mixed following
mergers. 
A scenario that incorporates this expectation 
and leads to 
the required observational outcome of $n_{EUVe} \propto n^2_{te}$ 
is that
the EUV 
emission is the result of secondary electrons and positrons 
(hereafter 
SEP) produced as byproducts of inelastic collisions between well-mixed 
primary cosmic ray protons and the thermal intracluster plasma.
The production rate
for SEP scales as 
$n_{cp} \times n_{tp}$, where $n_{cp}$ is the
density of cosmic ray 
protons. Then if the SEP energy losses are
independent of cluster 
position, as they would be for IC-CMB-dominated
losses, the consequent 
distribution of EUVe is the required $n_{EUVe} \sim
n_{SEP} \propto 
n^2_{tp} \sim n^2_{te}$.

A variety of primary cosmic ray source mechanisms
such as supernovae, active galaxies, and
terminal galactic wind shocks
might lead to the 
needed spatial distribution for this explanation of
the EUV excess, so 
long as their contributions were spread over enough
time and were spatially distributed into the 
various components.

However, 
another source seems to us to be the most 
likely candidate for these particles: cosmic rays accelerated at the
large scale 
``structure shocks" that accompany cluster formation.
These include what 
are usually termed merger and accretion shocks,
although recent 
cosmology simulations demonstrate a more complex and
richer shock 
pattern than those labels suggest (Miniati \etal 2000; Ryu \etal 2003). 
In a
cosmic structure 
simulation that included shock-accelerated cosmic
rays Miniati \etal 
2001a indeed found that the cosmic ray proton
distribution in the 
central regions of their clusters scaled
roughly with the thermal gas
although the two distributions did show differences from cosmic rays associated
with recent shocks, especially outside the cluster core region.

A more detailed evaluation of our model requires that we estimate the 
population 
of cosmic rays responsible for the SEP.
Inelastic collisions between cosmic ray protons with kinetic
energies above about 300 MeV and
the thermal intracluster medium produce mainly charged and neutral pions. The charged
pions decay into muons and neutrinos, and the muons into the SEP
that are responsible for the EUV emission. 
We express the cosmic ray proton 
density distribution as a power-law of the form
\begin{equation}
n(\gamma_p) = n_{p0} \gamma_p^{-s_p},
\label{pspec}
\end{equation}
where $E_p = \gamma_p M_p c^2$ is the proton energy.

The
approximate formalism given in Mannheim and Schlickeiser 1994
gives us a simple expression for the
net omnidirectional SEP production rate, $q_e$, with 
energies 
$\gamma m_ec^2 >> 35$ MeV from p-p collisions. In particular, for a 
power law proton energy distribution we have
\begin{equation}
q_e \approx \frac{13}{12} \sigma_{pp} c n_{tp} n_{p0} 
\left(\frac{M_p}{24 m_e}\right)^{s_{e0}-1} 
\gamma^{-s_{e0}}~~{\rm cm}^{-3}{\rm s}^{-1},
\label{sep_rate}
\end{equation}
where $\sigma_{pp} \approx 3.2 \times 10^{-26} {\rm cm}^{-2}$,
$s_{e0} = \frac{4}{3}(s_p - \frac{1}{2})$, and $q_e^+ \approx q_e^-$.
At low energies this expression overestimates SEP production, so
we have compared it with a numerical calculation for $q_e^{\pm}$
based on the more accurate pion production described in association
with equation \ref{gamma-emiss} (following) and the SEP distribution given by
Moskalenko \& Strong 1998. For electron energies of 150 MeV the two 
results agree to better than about 50\% 
so in what follows we use the simpler expression 
in equation \ref{sep_rate}.
We argued above that the lifetimes of the EUVe at $\sim 200$ MeV SEP
are determined by their energy losses against 
IC-CMB. We can reasonably assume that the SEP density
is set by a balance between p-p production and IC-CMB losses. Since the
EUV emission is the same IC-CMB, it is straightforward to derive
the expected omnidirectional EUV volume emissivity directly
in terms of the cosmic ray proton density. The result using equation
\ref{sep_rate} is
\begin{equation}
\nu_i \epsilon_{\nu_i}  \approx j^{BC}_{\alpha o} \frac{26}{12(s_p - 
5/4)}
\left(\frac{M_p}{24 m_e}\right)^{s_e - 2} \times \\
\label{sep-ic}
\end{equation}
\[
\sigma_{pp} c n_{tp} n_{p0}
m_e c^2 \left(\frac{\nu_{\mu}}{\nu_i}\right)^{\alpha - 1}~~{\rm 
erg~cm}^{-3}~{\rm s}^{-1},
\]
where $\nu_i$, $\nu_{\mu}$ and $j^{BC}_{\alpha o}$ were identified in 
relation to 
equation \ref{synch-ic}, $s_e = \frac{4}{3}(s_p + \frac{1}{4})
= s_{e0}+1$ is
the spectral index of the steady-state SEP energy distibution,
and, once again, $\alpha = (s_e - 1)/2$. 
With a constant cosmic ray density fraction, $f_p$ =
$n_{p0}/n_{tp}$,  the EUV emissivity
scales with $n_{tp}^2$, as required by the EUVE data for Coma. 

Using this result we can integrate over the cluster to compute
an EUV luminosity, which can then be compared with the observational
result. The result will depend on an assumed cosmic ray
energy spectral index as well as on the intracluster medium density distribution.
If the cosmic ray proton flux is due to structure
formation shocks, the cosmic 
ray 
spectrum represents
an average from the shocks dissipated in the local gas over cosmic
time. In the test particle limit for diffusive shock acceleration, 
which is a reasonable
approximation for relatively weak shocks, the standard relation
is $s_p = 2(M^2 + 1)/(M^2 - 1)$, where $M$ is the shock Mach number.
Strictly speaking, this index applies to the momentum
spectrum, $n(p_p)$, of cosmic ray protons accelerated at
shocks, rather than the energy
spectrum, $n(\gamma_p)$, that we defined in equation \ref{pspec}
in order to apply analytic expressions for our simple model estimates. 
The two spectra compare as $n(\gamma_p)/n(p_p) = (1 - 1/\gamma^2_p)^{s_p/2}$.
At relativistic energies the two forms converge; at the threshold for
pion production, $\gamma_p \approx 1.3$, $n(\gamma_p)/n(p_p) \approx 1/3$,
which roughly compensates for the overestimate in SEP production 
from equation \ref{sep_rate}.

In a recent detailed analysis of shocks formed in a high resolution
cosmic structure formation simulation, Ryu \etal 2003 found that
the most important shocks for cosmic ray acceleration were
those with $M\sim 2-4$, corresponding to $s_p \sim 2.3 - 3$. To be 
specific in our estimates below 
we choose 
$s_p = 2.5$, 
which leads to $s_e = 11/3$ and $\alpha = 4/3$ which will be
consistent with the radio halo and HXR constraints. 
Then assuming
a beta law distribution for the intracluster medium, $n_{tp}(r) = 
n_{t0}/(1+(r/a)^2)^{3\beta/2}$,
with $n_{t0} = 3\times 10^{-3}~{\rm cm}^{-3}$, $\beta = 0.75$, (Briel 
\etal
1992)
and $a = 300$ kpc (corresponding to 10\arcmin.5 at 100 Mpc),
we compute the spatially integrated EUV flux to be
$\nu_i F_{\nu_i} \approx 1 \times 10^{-4} f_p~{\rm erg~cm}^{-2}{\rm 
s}^{-1}$.
Comparing this result to the observed $\nu_i F_{\nu_i} \approx 1.4\times 
10^{-11}~ {\rm erg~cm}^{-2}{\rm s}^{-1}$, 
we obtain
$f_p \approx 1.4\times 10^{-7}$.

A test of the reasonableness of this result can be made by determining 
its 
consistency with upper limits on the $\gamma$-ray flux in the Coma 
cluster. In
addition to charged pions, inelastic p-p collisions will produce 
neutral
pions, which will quickly decay to $\gamma$-rays. The resultant 
$\gamma$-ray
spectrum peaks near 70 MeV, but extends to higher energies and,
in particular, into the $\ge 100$ MeV EGRET band. 

The $\gamma$-ray emissivity due to a power-law cosmic ray spectrum
approaches a power-law at high
energies, making it relatively straightforward to compute an analytical
estimate of the high energy flux. 
However, the EGRET band is too close to the 70 MeV peak
for that approximation to be adequate for our needs. Fortunately,
semi-empirical relations for our range of interest are 
available in the literature.
We have followed the formulation laid out
conveniently in Schlickeiser 2002. The omnidirectional $\gamma$-ray
emissivity can be written as
\begin{equation}
q_{\gamma}(E_{\gamma}) = 2 
\int^{\infty}_{E_{\gamma}+(m_{\pi}c^2)^2/(4E_{\gamma})}
\frac{q_{\pi^0}(E_{\pi})}{\sqrt{E^2_{\pi} - (m_{\pi}c^2)^2}}dE_{\pi},
\label{gamma-emiss}
\end{equation}
where $q_{\pi^0}(E_{\pi}) \propto \sigma_{pp}cn_{tp}n_{p0}$ is
the neutral pion production rate. That rate asymptotes 
to a power-law at high energies with spectral index $s_{e0}$,
just as for the SEP, but drops sharply as the pions become 
nonrelativistic.
Using the full expressions given by Schlickeiser, 
assuming $s_p = 2.5$ and integrating equation \ref{gamma-emiss} over
photon energy, we obtain 
$q_{\gamma}(E_{\gamma}\ge 100 {\rm MeV})$
$\approx 0.32 \sigma_{pp}cn_{te}n_{p0}~{\rm ~photons~cm}^{-3}~{\rm 
s}^{-1}$.  From the values established above for Coma, including the required 
cosmic ray density fraction, $f_p$, we obtain an estimated
$\gamma$-ray flux $F_{\gamma}(E_{\gamma}\ge 100 {\rm MeV}) \approx
1.4\times 10^{-9}~{\rm ~photons~cm}^{-2}~{\rm s}^{-1}$. 
Sreekumar \etal 
1996 give a 2 $\sigma$ upper flux limit in this band 
of $4\times 10^{-8}~{\rm photons~cm}^{-2}~{\rm s}^{-1}$ for the Coma 
cluster.
This is well 
above
the $\gamma$-ray flux produced by our required SEP population.

Simulations
such as those of Miniati \etal 2000, 2001a,b and Ryu \etal 2003
have suggested that structure formation shocks might lead to
cosmic ray energy pressures approaching as much as 1/3 the total
intracluster medium pressure.
Accordingly, we have estimated the cosmic ray energy density in
the core of Coma that would be necessary under the SEP
model we have proposed. For the proton
spectra of immediate interest ($s_p \sim 2.5$), most of the
kinetic energy resides in mildly relativistic particles, independent of
whether we use the energy power law of equation \ref{pspec} or
the analogous momentum power law. We can write approximately
$u_{cp} \sim n_{p0} M_pc^2$, which gives $u_{cp} \sim 10^{-12}{\rm erg~cm}^{-3}$
for the nonrelativistic cosmic ray proton distribution. This compares
to the thermal energy density, $3n_{tp}kT \approx 4\times 10^{-11}{\rm erg~cm}^{-3}$.
This rough estimate can be compared to estimates from cosmology
simulations which range upwards of tens of percent
(e.g. Miniati \etal 2001a; Ryu \etal 2003).

We can summarize the constraints on our model as follows. 
The electron energy spectrum must have a power law
slope steeper than $s_e \approx 3.2$, in order to avoid excess inverse-Compton
hard X-ray emission. Constraints set by the observed radio
synchrotron flux depend on the spectral index of the radio emission and the 
cluster magnetic field.
An electron spectrum with $s_e = 3$, requires the cluster magnetic field
be $\sim 0.1 \mu$Gauss.
If $s_e = 4$ the cluster field must be $\sim 1.4 \mu$Gauss.
These outcomes are shown graphically in Figure \ref{b-field}.
\begin{figure}
\begin{center}
\plotone{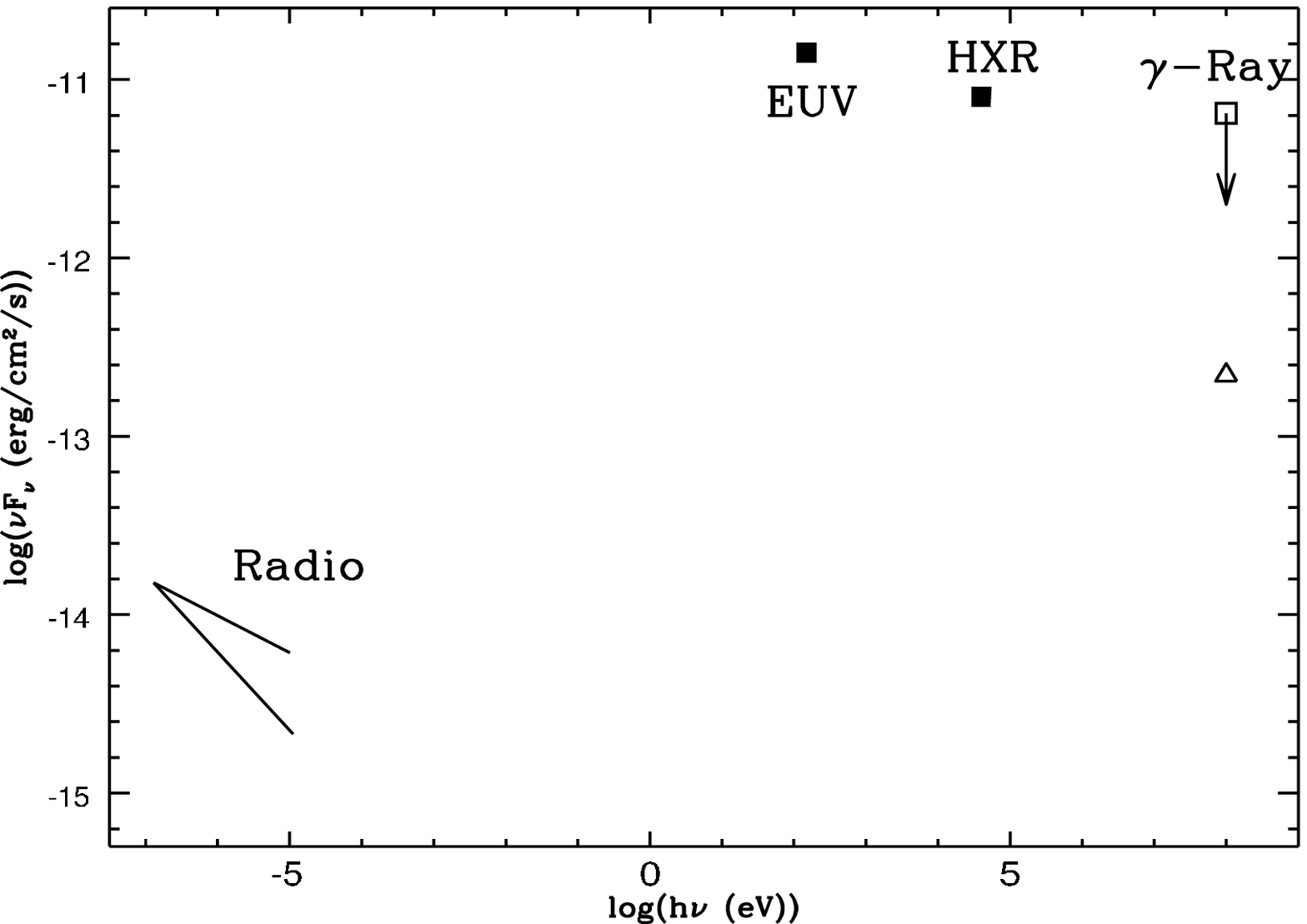}
\caption{
\captionsix
\label{b-field}}
\end{center}
\end{figure}

\section{Conclusions}

We have analyzed archival data obtained with EUVE on the core of the 
Coma 
cluster. We find the ratio of the azimuthally averaged EUV and X-ray 
intensities is 
essentially constant with increasing cluster radius. In addition, a 
correlation of the 
diffuse EUV emission with the diffuse X-ray emission shows that the 
detailed 
spatial 
distributions of these emissions are quite similar, but not identical. 

XMM-{\em Newton} and {\em Chandra} observations 
show that there is no 
intracluster 10$^6$ K gas in the core of the cluster and hence that the EUV 
emission must 
be non-thermal. 
The only viable non-thermal source for the EUV excess is photons inverse 
Compton scattered by 200 MeV electrons from the cosmic 
microwave 
background. To account for the observed EUV intensity distribution, the 
scattering electrons must be distributed with a spatial density roughly 
in proportion to the square of the thermal plasma density. A scenario 
that naturally produces these results is that the EUV emitting 
particles are electron/positron secondaries produced by 
inelastic collisions between primary cosmic ray protons and thermal 
protons with similar spatial distributions. This type of distribution would be a natural outcome if the cosmic 
rays had been produced over the past several Gyr and had become well-mixed with the intracluster 
medium.

This model accounts naturally for the average spatial distribution of the 
EUV emission in relation to the X-ray emission. It accounts for the 
similar, but 
imperfect, pixel-to-pixel correspondence of the EUV and X-ray emission.
Finally, it accomplishes this without 
violating 
observational limits in other bands of the spectrum.  We have demonstrated that
the required underlying cosmic rays could reasonably have been produced in large
scale structure shocks accompanying the cluster formation.

Secondary electrons as the source of the emission 
in radio halos were first suggested by
Dennison 1980 and have since been discussed by a large number of 
authors. However, observational 
evidence for the presence of these particles has been lacking.  
There is, in fact, a growing body of evidence that, {\em in general},  
these particles may not be the 
underlying source 
of the emission in radio halos (Brunetti 2003; Kuo et al. 2003). 
Hence the EUV emission in the Coma 
cluster may 
be the only direct evidence for secondary electrons in an intracluster 
medium.

\acknowledgements{
We thank Pat Henry, Gianfranco Brunetti and Vahe Petrosian for extremely useful 
discussions.  
This work has been supported in part by a 
University of California Faculty Research 
Grant to S. Bowyer. M. Lampton acknowledges the support of 
the Director, Office of Science of the U.S. Department of Energy under
contract number DE-AC03-76SF00098.  
E.~J. Korpela's work is supported by NASA through grant NAG5-12424 and 
the NSF
through grant AST03-07956.
T.~W. Jones has been supported in this work by the NSF through grants
AST00-71167 and AST03-07600, by NASA through grant NAG5-10774
and by the University of Minnesota Supercomputing Institute.
}

\appendix
{\em Note added in proof.-} Miniati has recently remodeled the nonthermal
emission from clusters with an improved treatment (F. Miniati, MNRAS, 342, 1009
[2003]). In his published work, he shows results only above 10 keV.  However, he
has now extended his work to lower energies and has calculated the EUV flux with
the specific parameters of the Coma Cluster (F. Miniati, 2004, private
communication).  He finds that the IC EUV emission from SEPs is consistent at
the 20\% nominal level with the measured EUV flux reported here.  Further the
spatial distribution for this emission is concentrated in the inner Mpc of the
cluster.  These results provide additional support for the ideas presented
here.

\ifms{
\clearpage

\begin{figure}
\begin{center}
\plottwo{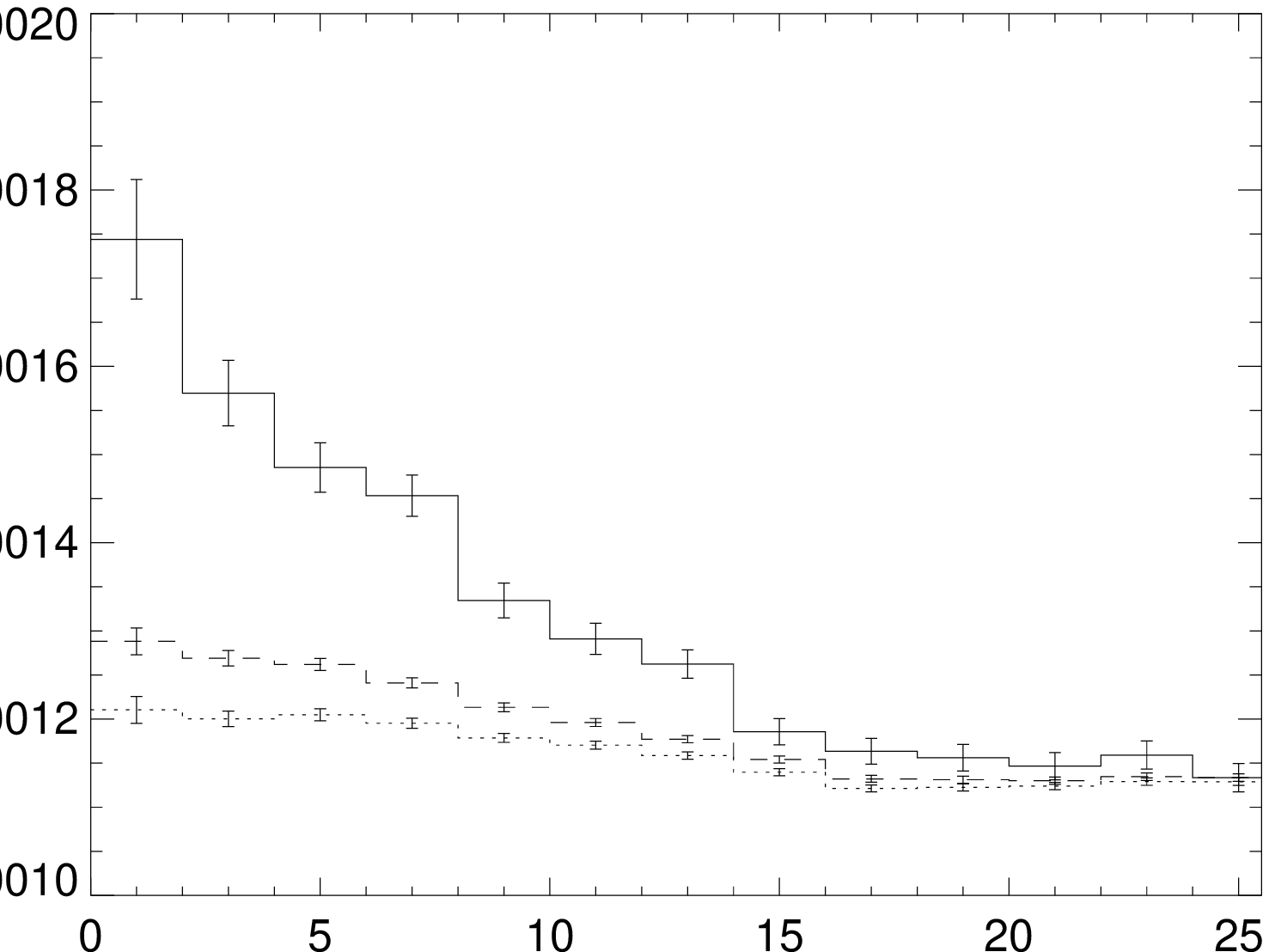}{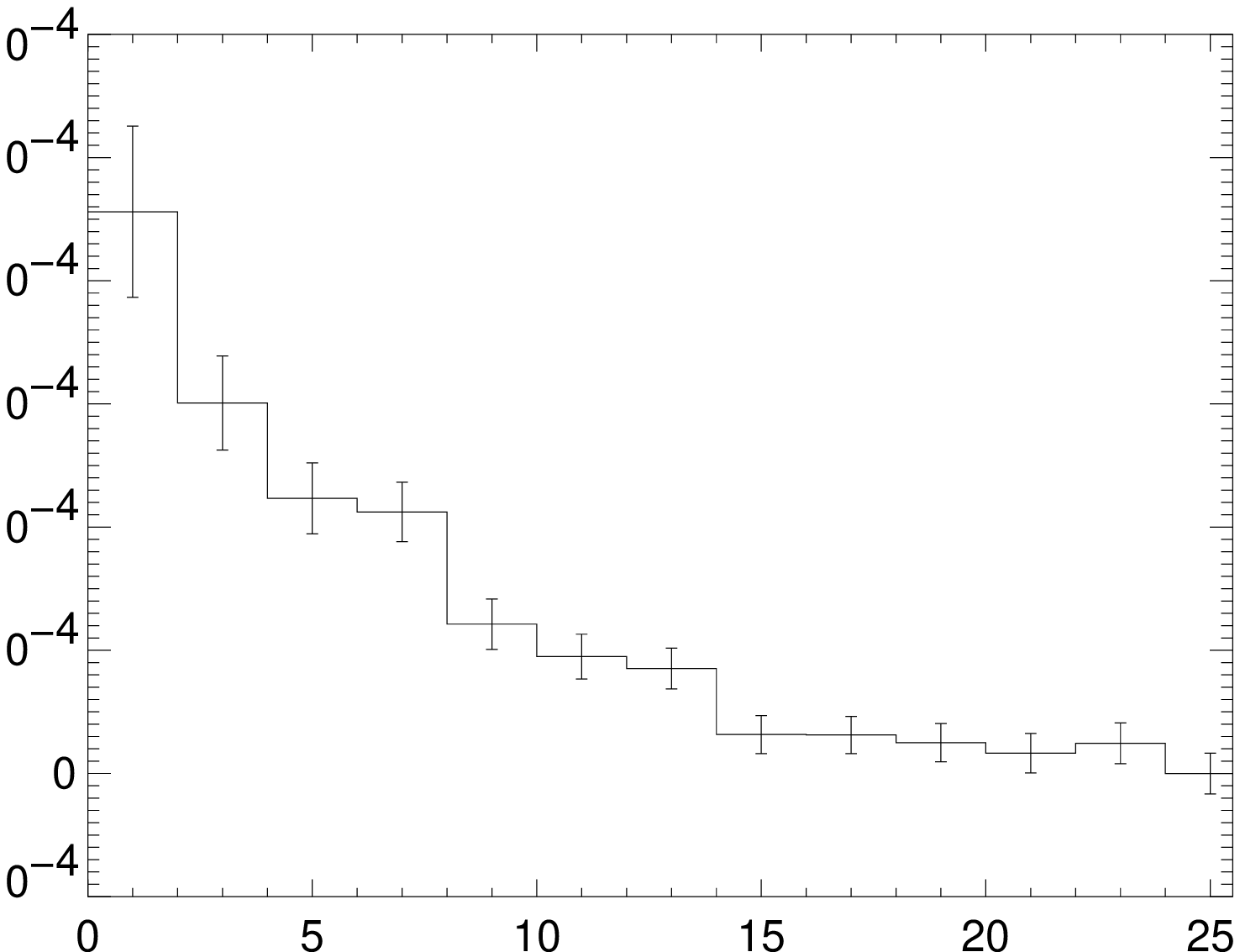}
\caption{
\captionone
\label{radial}}
\end{center}
\end{figure}

\clearpage
\begin{figure}
\begin{center}
\plotone{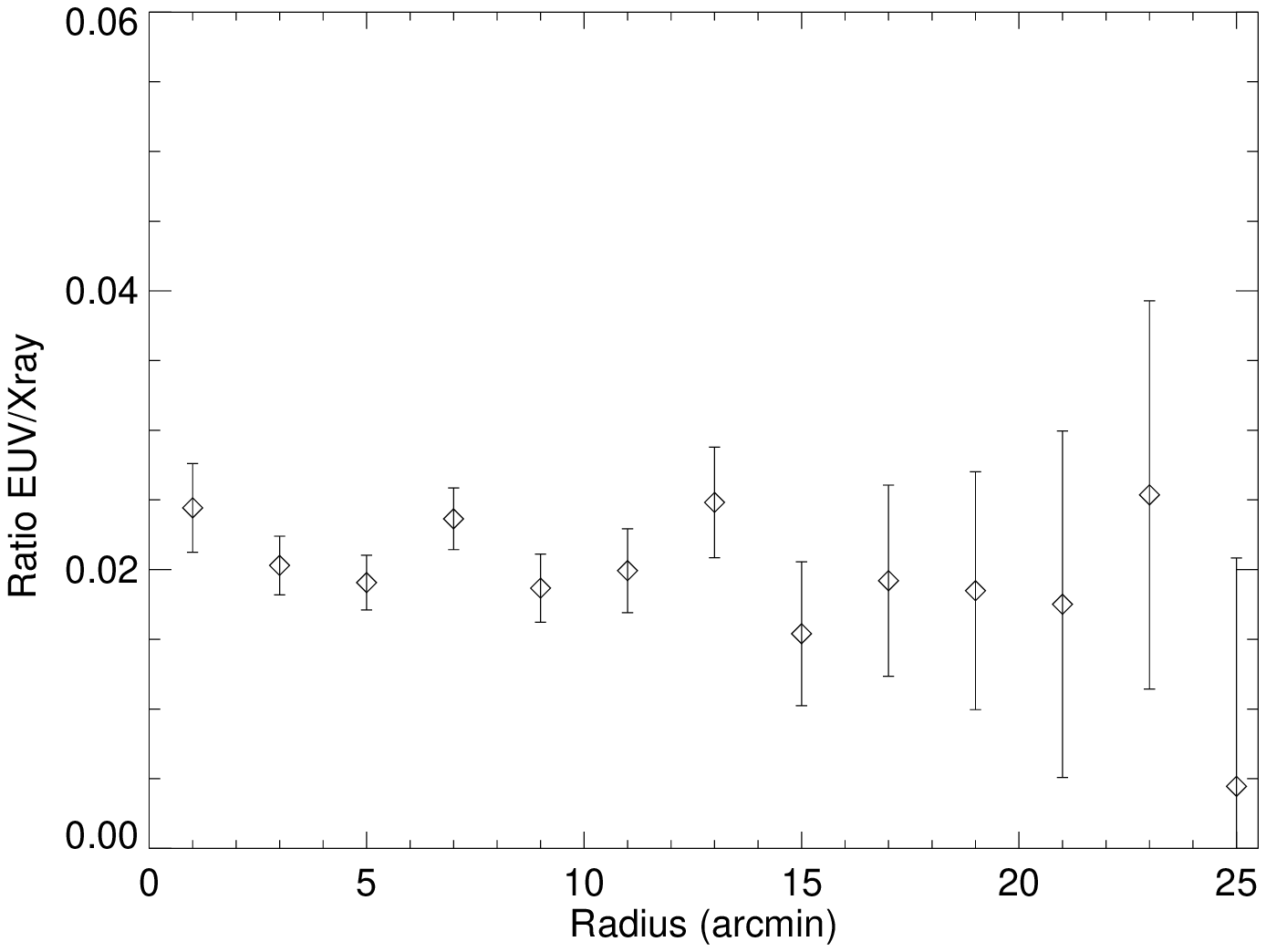}
\caption{
\captiontwo
\label{ratio}}
\end{center}
\end{figure}

\clearpage
\begin{figure}
\begin{center}
\plotone{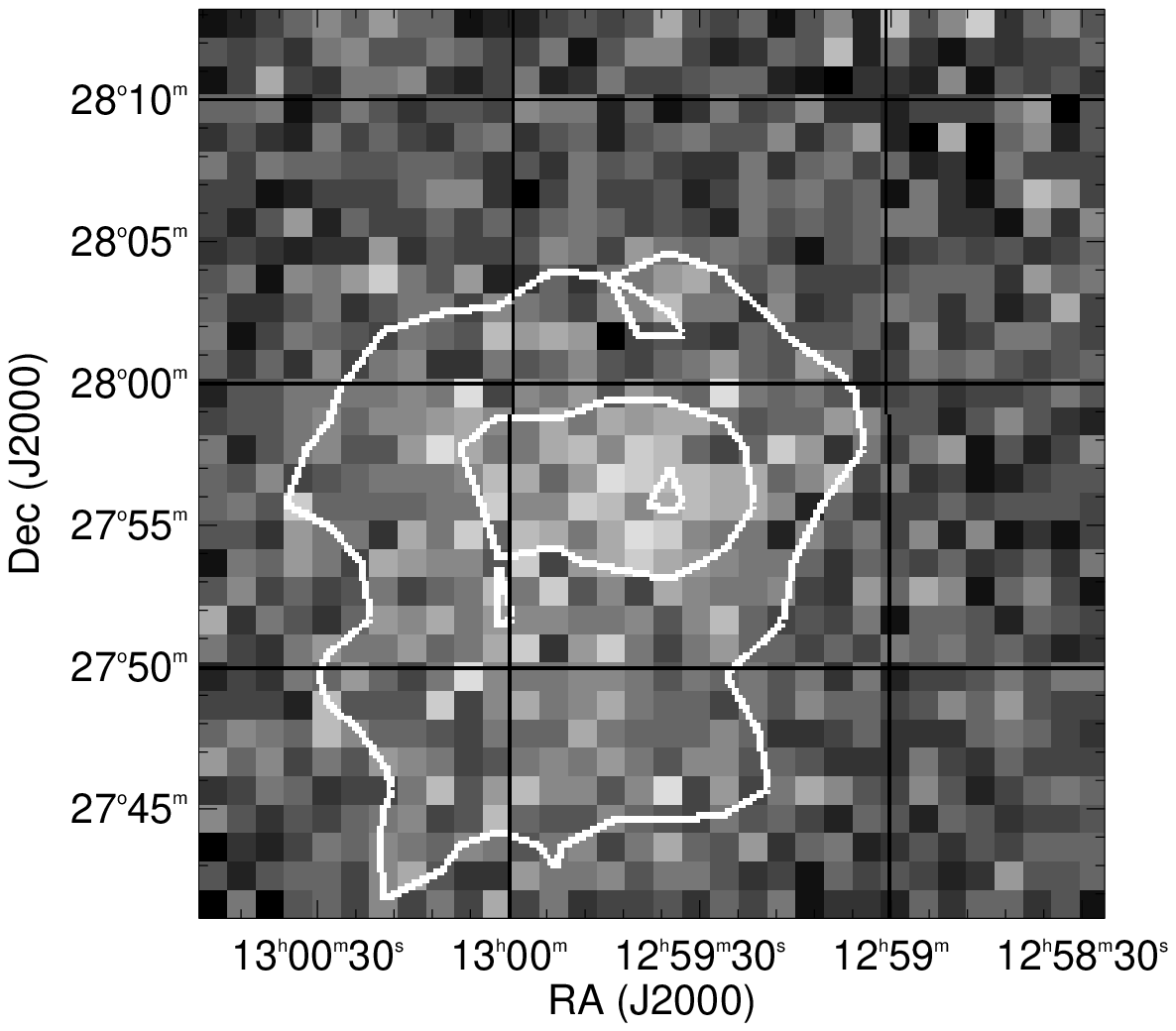}
\caption{
\captionthree
\label{map}}
\end{center}
\end{figure}

\clearpage
\begin{figure}
\begin{center}
\plotone{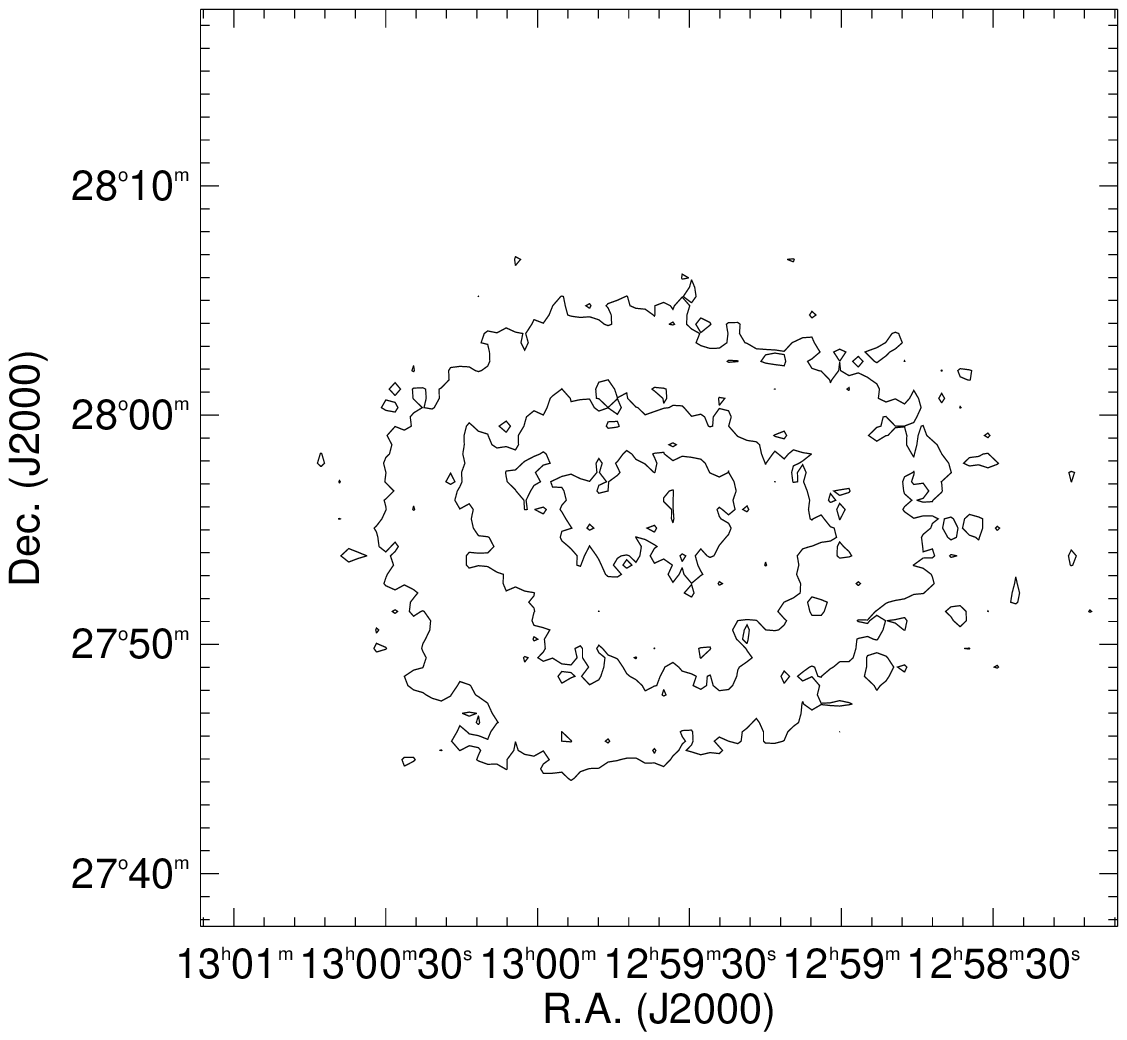}
\caption{
\captionfour
\label{xray}}
\end{center}
\end{figure}

\clearpage
\begin{figure}
\begin{center}
\plotone{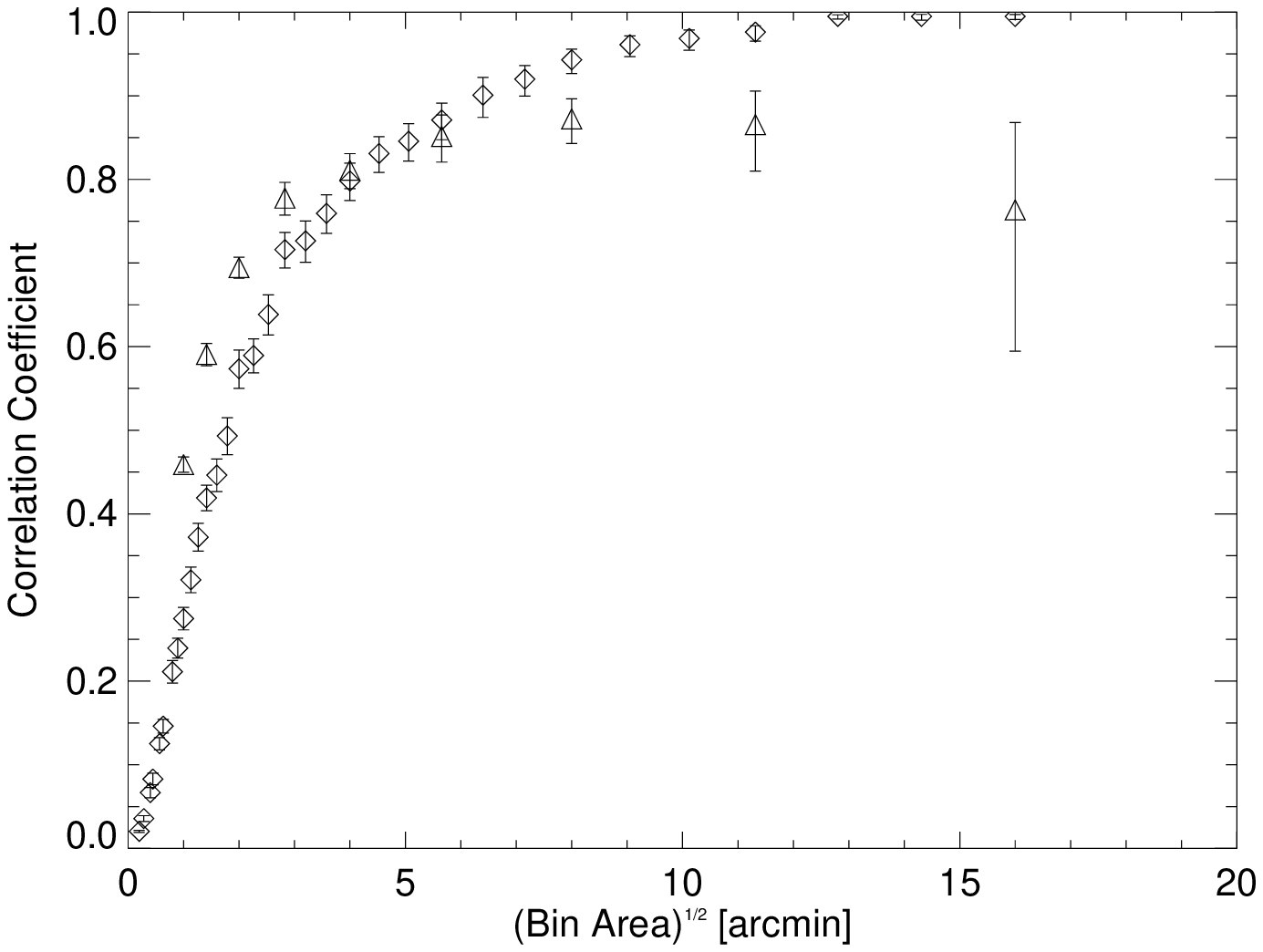}
\caption{
\captionfive
\label{correlation}}
\end{center}
\end{figure}

\clearpage
\begin{figure}
\begin{center}
\plotone{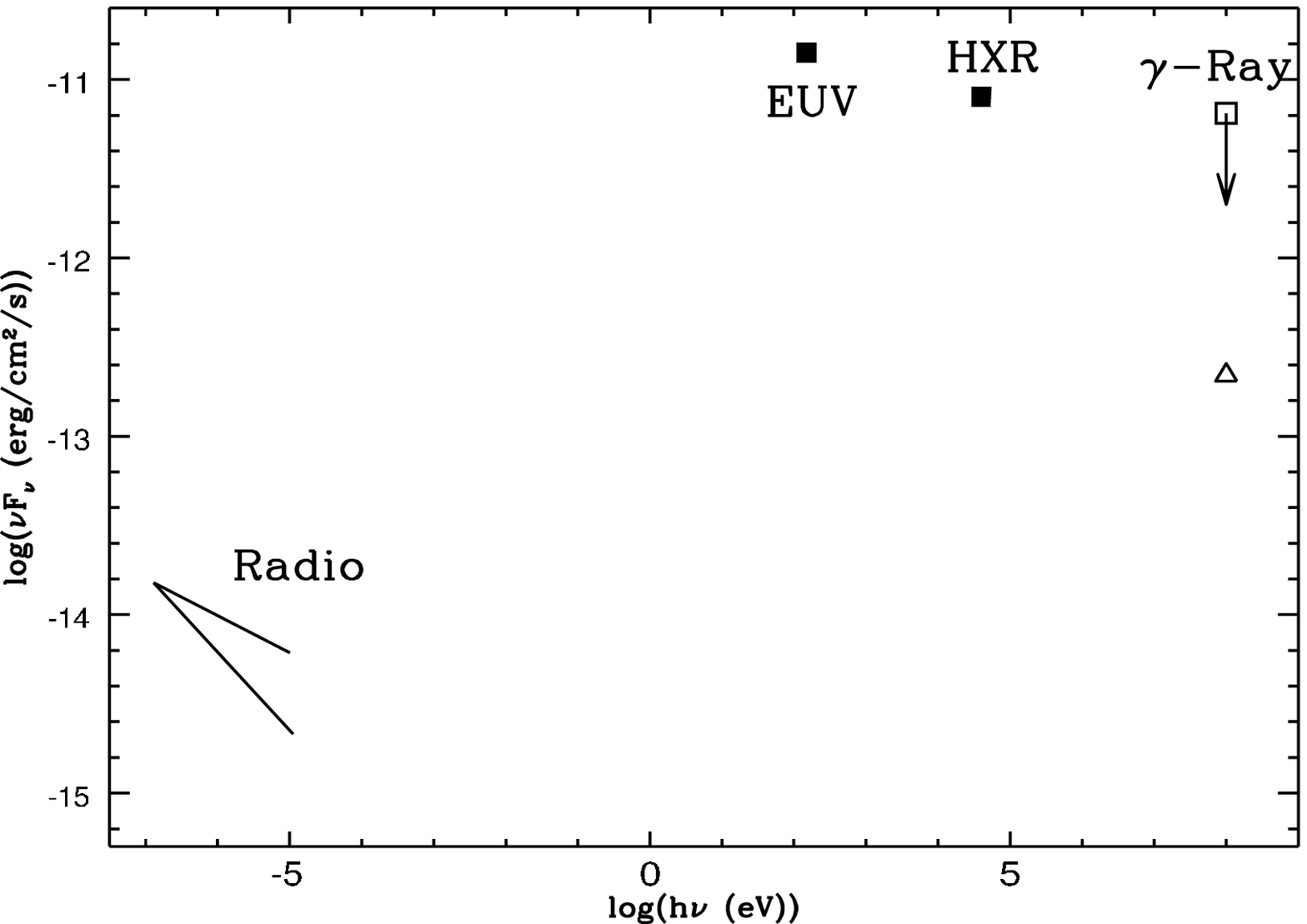}
\caption{
\captionsix
\label{b-field}}
\end{center}
\end{figure}

\clearpage
\begin{table}
\begin{center}
\begin{tabular}{lr}
\hline
Date    & \multicolumn{1}{c}{Duration (ks)} \\
\hline
\hline
12/25/95-12/28/95	&	50 \\
06/11/96-06/12/96	&	39 \\
01/12/99-01/14/99	&	53 \\
02/04/99-02/07/99	&	76 \\
03/15/99-03/21/99	&	172 \\
\hline
\end{tabular}
\caption{Log of Observations}
\end{center}
\end{table}

}
\end{document}